\documentclass[conference,a4paper]{IEEEtran}

\usepackage[utf8]{inputenc} 
\usepackage[T1]{fontenc}
\usepackage{url,hyperref}
\usepackage{ifthen}
\usepackage[noadjust]{cite}
\usepackage[cmex10]{amsmath} 

\usepackage{xparse,amssymb,mathtools,color,graphicx,algpseudocode,algorithm,comment,dsfont,subfigure,amssymb,stackengine,bm}

\newtheorem{theorem}{Theorem}
\newtheorem{lemma}{Lemma}
\newtheorem{proposition}{Proposition}
\newtheorem{corollary}{Corollary}

\newtheorem{definition}{Definition}

\newtheorem{remark}{Remark}

\newcommand{\Z}{\mathbb{Z}}

\newcommand{\C}{\mathbb{C}}

\NewDocumentCommand\sqn{mg}{%
    \|\mathbf{#1}_{\IfNoValueTF{#2}{}{#2}}\|^2%
}

\def\bs{\pmb}
\def\b{\mathbf}

\graphicspath{{../../Figures/}} 
\DeclareGraphicsExtensions{.pdf,.jpeg,.png,.jpg}

\DeclareMathOperator*{\argmax}{arg\,max}

\DeclareMathOperator{\EX}{\mathbb{E}}

\begin{document}
\title{Multi-Antenna Jamming in Covert Communication} 

\author{\IEEEauthorblockN{Ori Shmuel, Asaf Cohen, Omer Gurewitz}
\IEEEauthorblockA{Ben-Gurion University of the Negev, \{shmuelor, coasaf, gurewitz\}@bgu.ac.il
}
}

\maketitle

\begin{abstract}
Covert communication conceals transmission of messages from Alice to Bob out of a watchful adversary, Willie, who tries to determine if a transmission took place or not. While covert communication in a basic, vanilla setting where all variables are known to Willie, results in the well-known square-root law, when a jammer is present and assists Alice by creating uncertainty in Willie's decoder, a strictly positive transmission rate is possible. 

In this work, we analyze the case where the jammer is equipped with multiple antennas. Specifically, we analyze the effect of multiple antennas at the jammer on Alice's transmission power and consequently on the transmission rate. We consider both cases, one in which the channel knowledge is known and one in which it is unknown by the jammer. We formulate several optimization problems for the transmission strategies of the jammer, to maximize his assistance to Alice, in terms of maximizing a ratio between Willie's and Bob's noise variances. 

When the channel information is known to the jammer, we show that the optimal strategy of the jammer is to perform beamforming towards a single direction with all his available power. This direction though, is not trivial, since it reflects an optimal tradeoff point between minimizing the interference at Bob and maximizing the interference at Willie. When the channel knowledge is unknown, we show that the optimal strategy of the jammer is either to transmit isotropically to all directions or to the null-space of Bob, where this choice depends on certain channel conditions. This is in contrast to current schemes in the literature. Furthermore, we extend the optimization problems to the case where Bob is also equipped with multiple antennas, and provide insightful results, shown to be asymptotically optimal, accompanied by simulations. 
\end{abstract}

\section{Introduction}
In covert communication (also known as Low Probability of Detection - LPD) Alice tries to reliably communicate a message to Bob, such that a watchful adversary Willie remains unaware of the presence of the communication. To make this possible, Alice may use the fact that the channels between all participants are subject to some kind of noise, and therefore she can try to hide her communication within the margin of uncertainty at Willie's decoder. In fact, for AWGN channels, it was shown in \cite{bash2013limits} that Alice can covertly transmit $O(\sqrt{n})$ bits in $n$ channel uses (a.k.a the square root law). Extensions for binary symmetric, discrete memoryless, multiple access and fading channels were done in \cite{che2013reliable,wang2016fundamental,bloch2016covert,arumugam2016keyless,tahmasbi2019covert}, respectively.  

This law essentially means that the transmission rate goes to zero with $n$; however, subsequent works showed that $O(n)$ bits in $n$ channel uses can be achieved, namely, a strictly positive rate, if Willie suffers from some kind of uncertainty in his received noise power \cite{che2014reliable,lee2015achieving,sobers2017covert}. The uncertainty may be a result of inaccurate knowledge of his noise or a result of an active node which confuses Willie (e.g., a jammer that varies his noise power randomly). The ability to achieve a strictly positive rate is of great importance since not only Alice can transmit a meaningful amount, but existing coding schemes can be used instead of designing special codes which are suitable only for covert communication. 

The limits of covert communication in a multiple-antenna setting were first established in \cite{abdelaziz2017fundamental}. Therein, it was shown that in case Alice is equipped with multiple antennas, her best strategy is to perform beamforming towards Bob, which results with a multiplicative constant gain to the square root law, by the number of independent paths between her and Bob. However, the case where such a communication channel includes a jammer which is equipped with multiple antennas is still open and remains unclear under various settings. For example, the knowledge the jammer has on the Channel State Information (in particular the CSI of Willie), and his preference on which user to assist, may affect his strategy, and the resulting rates, significantly.

In this work, we analyze the effect of multiple antennas at the jammer on covert communication, while assuming the jammer chooses to assist Alice and Bob. For simplicity, we assume that Alice and Willie have a single antenna. Note that a strictly positive power (hence, rate) can be achieved when there is a jammer with a single antenna (\cite{sobers2017covert}) and that the limits for MIMO settings without a jammer were already examined in \cite{abdelaziz2017fundamental}. Therefore, adding more antennas to Alice and Willie won't contribute much additional insight. 

The jammer's assistance comes in the form of transmitting Artificial Noise (AN) while using all his multiple antennas. Thus, his transmission strategy is reflected by the covariance matrix for the random vector he chooses. In a way, this covariance matrix defines the power allocation and the directions for that allocation. Accordingly, we analyze the behavior of Alice's transmission power as a function of the jammer's strategy which, to the best of our knowledge, was not yet examined under the settings of covert communication. Recently, and independently with this work, covert communication in the presence of a multi-antenna jammer was studied in \cite{forouzesh2020covert}. Therein, the authors \emph{fixed a certain jamming strategy}, of transmitting AN diffusely into the null-space of Bob, and optimized the covert rate by optimizing Alice's strategy (also having multiple antennas). In this work, we show that such a strategy is not always optimal, for both cases of known and unknown CSI of Willie, and show that under certain scenarios, transmitting AN outside the null-space of Bob, although it adds noise to his reception, is still better, since it is able to significantly jam Willie's receiver. 

The notion of AN transmission was also considered for wiretap channels, where the AN was used to improve the secrecy rate of the system. These works can be roughly divided into two cases, which differ by the source of the AN transmission. In the first case, it is assumed that the transmitter (Alice) is equipped with multiple antennas and she performs simultaneously the secret transmission and the AN. E.g.,  \cite{khisti2007secure,goel2008guaranteeing, mukherjee2009fixed, swindlehurst2009fixed, khisti2010secure, tsai2014power}. In the second case, there exist an additional node(s) in the system, e.g., a relay(s) \cite{dong2010improving, huang2011cooperative, wang2012distributed, luo2013uncoordinated} or a dedicated jammer \cite{fakoorian2011solutions, chen2015secrecy}, which output AN to degrade the eavesdropper's channel. At the heart of the analysis of these mentioned works, one can find several optimization problems depending on the models' assumptions. A major factor in all is the CSI, which is known or unknown to the nodes in the system. Regardless of this knowledge, it was shown in \cite{goel2008guaranteeing} that the transmission of AN can guarantee a minimal secrecy rate. This result is consistent with the fact that a positive rate is also achievable in covert communication problems when there exists uncertainty at Willie's detector.
Similarly to the wiretap channel problems, in this work, we are faced also with optimization problems for the optimal AN transmission strategy of the jammer. However, these optimization problems are originated from different settings, and therefore require different treatment due to the covertness requirement.

\subsubsection*{Main Contribution}
In this work, we explore the effect of multiple antennas of an uninformed jammer on covert communication. We analyze the covert rate as a function of the jammer's transmission strategy. We choose to use Bob's received SNR as our figure of merit since in most cases the actual rate of any coding scheme is an increasing function of the SNR. This is, of course, true for the AWGN channel as well. Specifically, we consider the optimization problems that arise when maximizing the received SNR at Bob by optimizing the jammer's strategy, defined by his covariance matrix. As part of the solution for these optimization problems, we concentrate on a covertness-achieving transmission scheme (construction) for Alice. Specifically, we provide the transmission power as a function of the jammer's transmission strategy for the cases where CSI on Willie is globally known or unknown. We show that this power ensures that the system is covert; that is, Willie has nothing better to do besides guessing if communication occurred or not. We show that multiple antennas at the jammer provide additional gain for Alice's transmission power with respect to the case of a single antenna jammer. Furthermore, when the CSI of Willie is unknown, we show that the transmission power of Alice is an increasing function of the number of antennas the jammer has.

Provided that the system is covert, i.e., assuming Alice uses the covertness achieving construction, we solve the optimization problem which turns out to be not trivial, as the jammer's strategy affects Bob's SNR in opposite directions. Nevertheless, under this construction, we solve both optimization problems for the cases of known and unknown CSI of Willie. The solutions essentially describe the directions and power allocations the jammer's transmission should use. When the CSI is known, the optimal transmission strategy for the jammer is to transmit in a single direction, with all of his available power. In general, this direction is not trivial, since it takes into account both the channel coefficients to Bob as well as those to Willie. Thus, the transmission direction has a tradeoff between being orthogonal to Bob and in the direction of Willie. When the CSI of Willie is unknown, yet the CSI of Bob is known to the jammer, we show that transmitting the AN diffusely into the null-space of Bob is not always optimal.

Furthermore, we consider the case where Bob is also equipped with multiple antennas and uses a linear receiver. This provides even further depth and complexity to the optimizations in question, as Bob takes an active part in the decoding process. We extend the optimization problems for this case and present several transmission strategies that the jammer should take to assist Alice as much as possible, as well as the detection strategies Bob should employ when he has multiple antennas. The suggested strategies are compared to the optimal solution using numerical results. For the low SNR regime, some of the suggested strategies are shown to be asymptotically optimal.

\section{System Model} \label{sec-System model}

\begin{figure}[t]
   \centering
   	\includegraphics[width=0.4\textwidth]{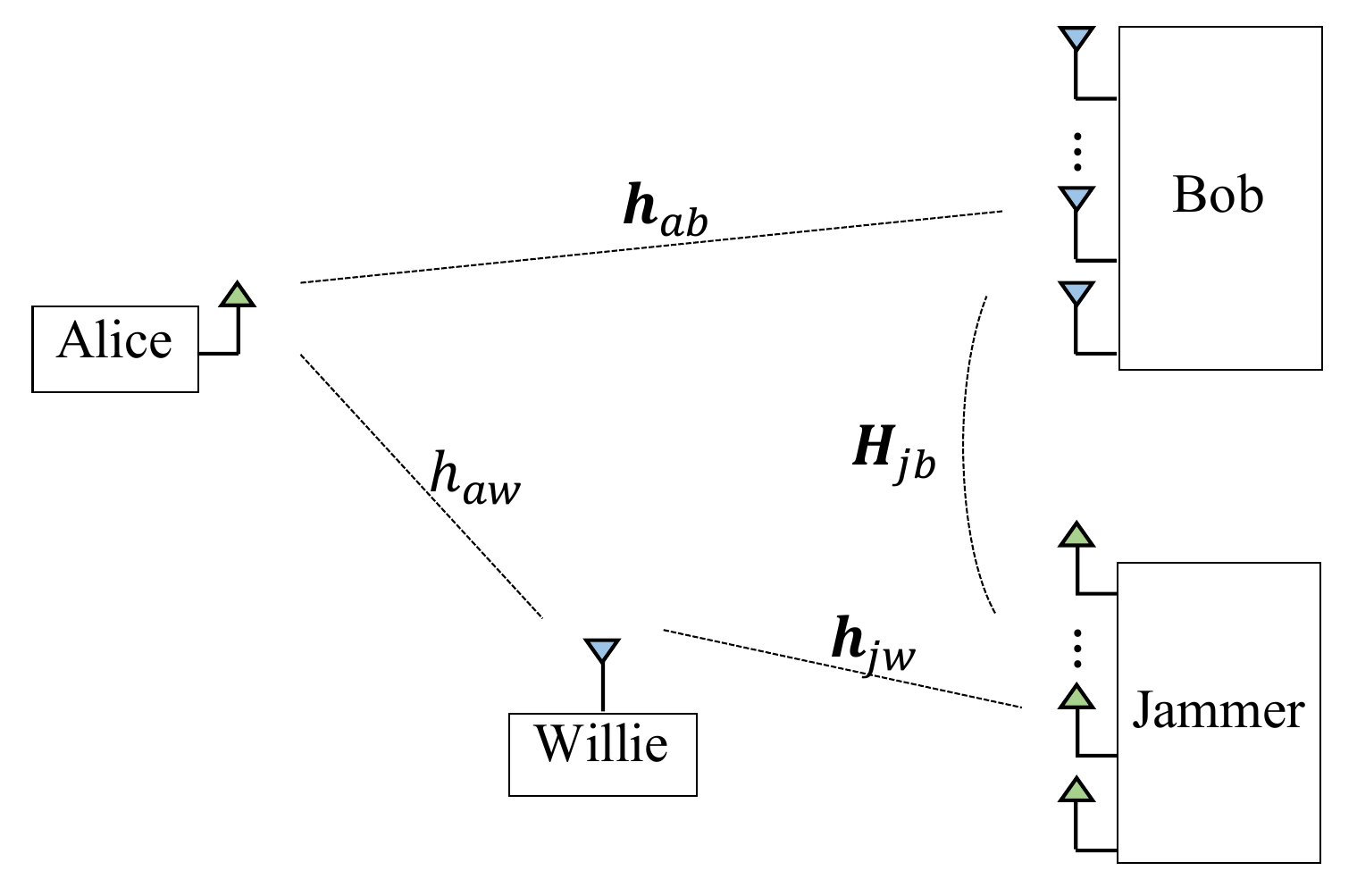}
   	\caption{Covert communication model where an independent jammer who assist the communication by transmitting AN.}
	 \label{fig-Model Bod is not jammer}
 \end{figure}

We consider a system in which Alice ("a" in the channel coefficients notation) wishes to communicate covertly with Bob ("b") while Willie ("w") remains uncertain of this communication. Also, we assume that there is a third participant, the jammer ("j"), who assists Alice and Bob.  
This paper focuses on covert communication with a positive covert rate. Thus, we assume that the uniformed jammer assists Alice in creating uncertainty at Willie's decoder continuously,
regardless of whether or not Alice transmits (Similarly to \cite{sobers2017covert}). The model is depicted in Figure \ref{fig-Model Bod is not jammer}.

Alice and Willie are equipped with a single antenna, while Bob and the jammer are equipped with $M$ and $N$ antennas, respectively. The channel between all participants is subject to block fading and AWGN. In this setting, Willie tries to detect whether transmission by Alice was made or not, by performing a statistical hypothesis test on his received signal. In this test, the null hypothesis $\mathcal{H}_0$ is that Alice does not transmit, while the hypothesis $\mathcal{H}_1$ is that Alice transmits. Throughout, lower case letters represent random variables, bold lower case represent random vectors, and bold upper case represent random matrices. Thus, under each of the hypotheses, the received signals at Bob and Willie in the $i$th channel use are
\begin{equation}\label{equ-The received signals at Bob and Willie}
\begin{aligned}
\mathcal{H}_1 :\quad &\b{y}_b[i]=x[i]\b{h}_{ab}+\b{H}_{jb}\b{u}[i]+\b{n}_b[i]\\
               \quad &y_w[i]=x[i]h_{aw}+\b{u}[i]^T\b{h}_{jw}+n_w[i] \\
\mathcal{H}_0 :\quad &\b{y}_b[i]=\b{H}_{jb}\sqrt{P_j}\b{u}[i]+\b{n}_b[i]\\
	       \quad &y_w[i]=\sqrt{P_j}\b{u}[i]^T\b{h}_{jw}+n_w[i],\\
\end{aligned}
\end{equation}
where $x[i]$ is the complex symbol transmitted by Alice in the $i$th channel use, with average power $P_a$ (i.e., $\EX[|x[i]|^2]=P_a$) and $\b{u}[i]=(w_1[i],w_2[i],...,w_N[i])^T$ is the vector transmitted by the jammer at the $i$th channel use, with a covariance matrix $\bs{\Sigma}=\EX[\b{u}[i]\b{u}[i]^\dagger]$ such that $\EX[\b{u}[i]^\dagger\b{u}[i]]=1$. $\b{u}[i]$ is multiplied by $\sqrt{P_j}$ where $P_j$ is the total transmission power of the jammer. 

The channel coefficients between Alice, Willie and Bob are $h_{aw}\in \C$ and $\b{h}_{ab}\in\C^{M}$, respectively. $\b{h}_{jw}\in\C^{N}$ and $\b{H}_{jb}\in\C^{M\times N}$ are the channel coefficients from the jammer to Willie and Bob, respectively. These channel coefficients are assumed to have a zero-mean complex Gaussian distribution with unit variance and are considered to be fixed for the period of $n$ channel uses (a slot). In addition, both Bob and Willie endure complex additive Gaussian noise denoted by $\b{n}_b\sim \mathcal{CN}(0,\sigma^2_b \mathbf{I}_{M\times M}) $ and $n_w\sim \mathcal{CN}(0,\sigma^2_w)$.

The CSI of Willie affects the ability of the jammer to contribute to the covert communication between Alice and Bob. We thus separate the scenarios where the CSI of Willie is globally known or unknown. When the CSI is known, we assume all nodes know $h_{aw}$ and $\b{h}_{jw}$. When the CSI is unknown, only the statistics of $h_{aw}$ and $\b{h}_{jw}$ are known to all nodes. In any case, we assume that Willie has full CSI of all channel coefficients and that Alice, Bob and the jammer know $\b{H}_{jb}$ and $\b{h}_{ab}$.



The jammer's assistance comes in the form of AN with a total power $P_j$ using his multiple antennas. Since he has $N$ antennas, he is free to choose how to allocate $P_j$ and to which direction to transmit. Hence, the transmitted vector by the jammer is $\b{u}[i] \sim \mathcal{CN}(0,\bs{\Sigma})$ where the total power $P_j$ is allocated according to $\bs{\Sigma}$. Again, such AN assistance does not impair the assumption that the jammer is uninformed. Note that the case in which the jammer has some kind of knowledge on when Alice is transmitting results in a very different problem, for which other type of assistance may be optimal.  

To create uncertainty at Willie's decoder, the jammer must vary his total power $P_j$, independently in each $n$ channel uses \cite{lee2015achieving,sobers2017covert,shahzad2018achieving}. That is, the value of $P_j$ must be unknown to Willie and changed in a way which he cannot estimate efficiently. Therefore, following similar assumptions as in \cite{sobers2017covert,shahzad2018achieving}, we assume that $P_j$ is a uniform r.v. on $[0,P_{max}]$ with the probability density function (pdf) given by,
\begin{equation}
     f_{P_j}(x) = 
  \begin{cases} 
   \frac{1}{P_{max}} & \text{if  } 0 \leq x \leq P_{max} \\
   0       & \text{otherwise }, 
  \end{cases}
\end{equation}
and it is redrawn every $n$ channel uses independently. The realization of $P_j$ in each slot is denoted by $p_j$. Throughout, we assume that $P_{max}$ is a fixed parameter of the system and it is known to all. Note also that if Willie had not known the channel coefficients, the jammer could have used a noise distribution with constant variance, since the uncertainty would arise from the random channel coefficients \cite{sobers2017covert}. 

As stated above, the jammer allocates its power $P_j$ in each slot according to the covariance matrix $\bs{\Sigma}$. His goal is to choose $\bs{\Sigma}$ which assists Alice and Bob as much as possible. Without loss of generality, the jammer transmits with $d\leq N$ directions with the fraction of power allocated to each direction represented by the vector $\bs{\xi}=(\xi_1,...,\xi_d)^T$ such that $\xi_l>0$ and $\sum_{l=1}^d \xi_l=1$. Thus, the covariance matrix $\bs{\Sigma}$ is represented as $\bs{\Sigma}=\b{V}\b{X}\b{V}^\dagger$ where $\b{X}$ is a diagonal matrix with $\bs{\xi}$ as its elements and $\b{V}$ is a matrix with the corresponding directions as its columns. Note that under this definition, the matrix $\b{X}$ is of dimension $d\times d$ and $\b{V}$ is of dimension $N \times d$.

Similar to previous works on covert communication (\cite{bash2013limits,lee2015achieving,sobers2017covert,shahzad2018achieving}), we assume that Alice and Bob share a codebook which is used only once and is not revealed to Willie ; however, Willie knows its statistics. 

\begin{remark}
Since the jammer assists Alice and Bob, this model and its analysis are also suitable, under some assumptions, to the situation where Bob is equipped with a full-duplex transceiver and he is the source of the AN. Moreover, in this case, any assumptions on the channel knowledge between them become trivial.  
\end{remark}

\subsection{Covert Criteria}
Upon receiving the vector $\b{y}_w$, Willie performs hypothesis testing to determine if transmission by Alice took place or not. That is, he tries to distinguish between the two hypotheses $\mathcal{H}_0$ and $\mathcal{H}_1$. The optimal test for that manner, in terms of minimizing the probability of error, is to apply the Neyman-Pearson criterion, resulting in the likelihood ratio test:
\begin{equation}\label{equ-likelihood ratio test}
\frac{\mathbb{P}_1}{\mathbb{P}_0} \Vectorstack{\mathcal{H}_1 > < \mathcal{H}_0} \eta,
\end{equation}  
where $\mathbb{P}_0$ and $\mathbb{P}_1$ are the probability distributions of Willie's observations under the hypotheses $\mathcal{H}_0$ and $\mathcal{H}_1$, respectively, and $\eta$ is a threshold which Willie can choose to trade-off between his two error probabilities, $P_{MD}$ and $P_{FA}$. $P_{MD}$ is the probability of miss detection in case a transmission occurred and $P_{FA}$ is the probability of false alarm in case a transmission did not occur.
In this work, we follow the standard criteria for covertness in the covert communication literature \cite{bash2013limits,lee2015achieving,bash2016covert,sobers2017covert}. Namely, we have the following. 
\begin{definition}[Covert Communication criteria]
\textit{Alice's transmission is considered covert if
\begin{equation}\label{equ-covert criteria}
P_{MD}+P_{FA}\geq 1-\epsilon,
\end{equation}
where $\epsilon >0$ is the covertness requirement.} 
\end{definition}
Throughout, we assume that $\epsilon$ is a fixed parameter of the system and it is known to all. Note that this criterion is acceptable for the following reason. Willie can easily choose a strategy with $P_{FA}=0$ and $P_{MD}=1$, by simply declaring $\mathcal{H}_0$ at all times, regardless of his channel measurements. Analogously, $P_{FA}=1$ and $P_{MD}=0$ are achieved by always declaring $\mathcal{H}_1$. Requiring $P_{MD}+P_{FA}\geq 1-\epsilon$ is therefore equivalent to forcing Willie to at most time-share between these two trivial strategies. 

\subsection{Performance Metric and Reliability}\label{sec-Performance Metric}

In this work, we examine the \emph{received SNR at Bob such that Alice's transmission is covert for a given covertness requirement $\epsilon$}. In what follows, we describe Bob's received SNR for fixed $\bs{\Sigma}, p_j$ and a fixed transmission power from Alice. First, for the case where Bob has a single antenna (i.e., $M=1$) and then for the case of multiple antennas which Bob may use beneficially. Such a distinction will help us focus only on the jammer's strategy first. Clearly, the SNR is affected by Alice's transmission power, regardless of the way it is chosen. In this work, we set this power such that the covertness constraint is met. Specifically, in Corollary \ref{cor-p_a as a function of the jammer's strategy only} given later on, we provide a sufficient condition for Alice's power, and prove that the resulting value is a function of solely $\bs{\Sigma}$, hence we write $P_a=P_a(\bs{\Sigma})$. Thus, in case Alice transmitted with some power $P_a(\bs{\Sigma})$, the received SNR of Bob is given as follows: 
\begin{equation}\label{equ-Bob SNR M=1} 
SNR_b^1= \frac{P_a(\bs{\Sigma})| h_{ab}|^2}{ p_j\b{h}_{jb}^\dagger \bs{\Sigma} \b{h}_{jb} +\sigma_b^2}=\frac{P_a(\b{V}, \b{X})| h_{ab}|^2}{ p_j\b{h}_{jb}^\dagger \b{V} \b{X} \b{V}^\dagger \b{h}_{jb} +\sigma_b^2},
\end{equation}
where $\b{H}_{jb}$, in this case, is a single column matrix denoted as $\b{h}_{jb}$. 

When Bob has multiple antennas, he can take an active part in the communication strategy. For example, by steering his antennas away from the jammer. We assume that Bob uses a linear receiver. In this case, Bob performs a linear operation on the received signal; specifically, Bob projects the received vector onto a subspace which on one hand diminishes the effect of the AN from the jammer and on the other intensifies Alice's transmission. 
That is, we have
\begin{equation*}\label{equ-The received signals at Bob after null steering}
\begin{aligned}
\b{c}^T\b{y}_b[i]&=x[i]\b{c}^T\b{h}_{ab}+\b{c}^T\b{H}_{jb}\sqrt{p_j}\b{u}[i]+\b{c}^T\b{n}_b[i],
\end{aligned}
\end{equation*}
where $\b{c}$ denotes the linear filter. The output of this receiver has SNR 
\begin{equation}\label{equ-The received SNR at Bob for a general projection c}
\begin{aligned}
\frac{P_a(\bs{\Sigma})|\b{c}^\dagger\b{h}_{ab}|^2}{\b{c}^\dagger\left(p_j\b{H}_{jb} \bs{\Sigma}\b{H}_{jb}^\dagger +\sigma_b^2\b{I} \right)\b{c}}=\frac{P_a(\b{V},\b{X})|\b{c}^\dagger\b{h}_{ab}|^2}{\b{c}^\dagger\left(p_j\b{H}_{jb} \b{V} \b{X} \b{V}^\dagger\b{H}_{jb}^\dagger +\sigma_b^2\b{I} \right)\b{c}}.
\end{aligned}
\end{equation}

The SNR is sufficient to derive the covert rate for the block fading and AWGN channels as it is an injective (and increasing) function of the SNR \cite{tse2005fundamentals}. However, considering the CSI knowledge assumptions taken in this work, the SNR expressions in \eqref{equ-Bob SNR M=1} and \eqref{equ-The received SNR at Bob for a general projection c} are random as they are a function of the varying AN noise variance $P_j$. Thus, to show that there exist a positive covert rate $R$ (without guaranteeing the maximum rate achievable) such that Bob can decode the transmission successfully with a probability of error that goes to zero, we note the following. First, $p_j$ is unknown to Bob and Alice; however, since the statistics of $P_j$ is known to all, Alice and Bob can derive a non-trivial bound as they know $P_{max}$. Second, when the CSI of Willie is known, the jammer's strategy, i.e., $\bs{\Sigma}$, can be computed directly from the specific realization of $h_{aw}$ and $\b{h}_{jw}$ and the bound on $p_j$. When the CSI of Willie is unknown, we chose to maximize a lower bound on the SNR, hence show that the jammer's strategy is characterized only by the rank used for $\bs{\Sigma}$. Specifically, under the assumptions made in this work, it is either $N-1$ or $N$, depending on the value of $p_j$. Accordingly, from the above, Alice and Bob can derive a lower bound on the SNR and agree on the covert rate such that the communication is reliable. 

We note that the SNR expressions above and the suggested analysis and results in this work can be used also for the scenario in which the channel coefficients $\b{h}_{ab}$ and $\b{H}_{jb}$ are unknown to Alice and Bob and only their statistics are known. In this case, there is a non-zero probability that the received SNR will be lower than any predefined value. For such fading channels, one of the acceptable performance metrics is the $\phi-$outage capacity. This is the maximal transmission rate $R$ such that the outage probability ${\text{P}_{\text{r}}}^{out}(R)$ is less than $\phi$ \cite[Ch. 5]{tse2005fundamentals}. Thus, for a given outage probability $\phi$, we say that Alice and Bob can communicate reliably and covertly with a covert rate $R$ and outage $\phi$, if the covert criteria \eqref{equ-covert criteria} is satisfied and $R<C_{\phi}$, where $C_{\phi}$ is the channel $\phi-$outage capacity from Alice to Bob. Now, to find such $R$, one needs to devise a different lower bound on the SNR by using the known distribution of the channel's fading coefficients along with the bound on $p_j$. For example, one can use a global lower bound for the SNR, which is a function of the channel's fading coefficients, given in Section \ref{sec-unknown CSI Willie M=1}, for that manner.

\section{Transmission Strategies and Alice's Transmission Power for $M=1$}\label{sec-AN Transmission Strategies M=1}

In this section, we analyze the jammer's AN transmission strategies along with the construction of a covertness achieving transmission scheme by Alice. Note that under our performance metric, Alice's transmission scheme is eventually measured by the power, $P_a(\bs{\Sigma})$, for a given $\epsilon$ and as a function of the jammer's strategy $\bs{\Sigma}$.

We start by formulating an optimization problem for the AN transmission strategy, to maximize Bob's received SNR, provided that the system is covert. We distinguish between the cases of known and unknown CSI of Willie, which influences the transmission strategy. As mentioned in Section \ref{sec-System model}, the Jammer's strategy is reflected by the covariance matrix $\bs{\Sigma}=\b{V} \b{X} \b{V}^\dagger$, which can be described by the matrices $\b{V}$ and $\b{X}$ alone since it assumed that the jammer uses all of his power, $p_j$, for transmission.   

Accordingly, we define the problem as follows 
\begin{equation}\label{equ-maximization problem of the jammer M=1}
\begin{aligned}
&\max_{\b{V,X}} \frac{P_a(\b{V},\b{X})| h_{ab}|^2}{p_j\b{h}_{jb}^\dagger \b{V} \b{X} \b{V}^\dagger \b{h}_{jb}+\sigma_b^2},\\
& s.t. \quad 0\leq\xi_l \leq 1,\quad \sum_{l=1}^N \xi_l =1,\\
&\ \quad\quad P_{MD}+P_{FA}\geq 1-\epsilon.
\end{aligned}
\end{equation}

\begin{proposition}\label{pro-optimal reduced optimization problem}
\textit{Let $P_a^{c,*}(\b{V},\b{X})$ be the optimal transmission power of Alice to assure covertness, i.e., satisfying \eqref{equ-covert criteria}. Thus, the optimization problem \eqref{equ-maximization problem of the jammer M=1} is equivalent to 
\begin{equation}\label{equ-maximization problem of the jammer M=1 sub}
\begin{aligned}
&\max_{\b{V,X}} \frac{P_a^{c,*}(\b{V},\b{X})| h_{ab}|^2}{p_j\b{h}_{jb}^\dagger \b{V} \b{X} \b{V}^\dagger \b{h}_{jb}+\sigma_b^2},\\
& s.t. \quad 0\leq\xi_l \leq 1,\quad \sum_{l=1}^N \xi_l =1.
\end{aligned}
\end{equation}}
\end{proposition}
Unfortunately, the optimization problem in \eqref{equ-maximization problem of the jammer M=1 sub} is non-convex and does not seem to have a closed form solution. This is mainly since $P_{MD}$ and $P_{FA}$ are affected by both the jammer's strategy $(\b{V},\b{X})$ as well as Alice's choice of power. Thus, the utility (SNR) and the constraint (covertness) interplay non-trivially.

Consequently, we suggest a sub-optimal solution by providing a transmission scheme by Alice which assures a sufficient condition for covertness, for every $\b{V}$ and $\b{X}$. This will allow us to provide an explicit expression for Alice's transmission power, ensuring that the system is covert as a function of $\b{V}$ and $\b{X}$. That is, we first make sure \eqref{equ-covert criteria} is satisfied and Willie is unable to decide if transmission occurred. We then find the optimal strategy by optimizing on $\b{V}$ and $\b{X}$. Thus, in the reminder of this section, \emph{an optimal strategy for the jammer refers to a one solving \eqref{equ-maximization problem of the jammer M=1 sub}, yet with a lower bound $P_a^c(\b{V},\b{X}) \leq P_a^{c,*}(\b{V},\b{X})$, which is given in the following lemma}. Specifically, Lemma \ref{lem-Alice power for covert system} bellow expresses Alice's transmission power, $P_a(\b{V},\b{X})$, as a function of any specific strategy, i.e., a fixed $\bs{\Sigma}$ and $p_j$ by the jammer, such that the communication is covert. 

\begin{lemma}\label{lem-Alice power for covert system} 
\emph{Assume a block fading AWGN channel and a jammer with $N$ antennas, who transmits AN with a fixed covariance matrix $\bs{\Sigma}$ and a total transmission power of $p_j$. For a given covertness requirement, $\epsilon >0$, as long as Alice transmits with power
\begin{equation}\label{equ-Alice power for covert system}
P_a\leq\frac{\epsilon P_{max}}{4|h_{aw}|^2}\b{h}_{jw}^\dagger \b{V} \b{X} \b{V}^\dagger \b{h}_{jw},
\end{equation}
the system is covert, i.e., \eqref{equ-covert criteria} applies and Willie is unable to decided if transmission occurred, and Alice's rate is strictly positive.}
\end{lemma}

\begin{IEEEproof}[Proof of Lemma \ref{lem-Alice power for covert system}]
\emph{Construction:}
We employ a Gaussian random codebook consisting of $M$ messages of size $n$. That is, the codebook is generated by independently drawing symbols from a zero-mean complex Gaussian distribution with variance $P_a$ and it is assumed to be used only once. When Alice wishes to transmit, she picks a codeword and transmits its $n$ symbols as the sequence $\{x[i]\}_{i=1}^n$. Such a construction was also used in other covert communication works for the AWGN channel such as, \cite{bash2013limits,sobers2017covert,shahzad2018achieving}. In what follows, under this construction, given $\epsilon$ and a fixed $\bs{\Sigma}$, we show how to set $P_a$ such that the covertness requirement \eqref{equ-covert criteria} is satisfied. Note that this construction will be used also in the proof of Lemma \ref{lem-Alice power for covert system with no CSI}.

\emph{Optimal hypothesis testing:}
The optimal test for Willie to distinguish between $\mathcal{H}_0$ and $\mathcal{H}_1$, given in \eqref{equ-likelihood ratio test}, can be written as
\begin{equation}
\frac{\mathbb{P}_1}{\mathbb{P}_0} =\frac{\prod_{i=1}^{n}f^1_{y_w[i]}}{\prod_{i=1}^{n}f^0_{y_w[i]}} \Vectorstack{\mathcal{H}_1 > < \mathcal{H}_0} \eta,
\end{equation}
where $f^0_{y_w[i]}$ and $f^1_{y_w[i]}$ are the probability distributions of Willie's observation in a single channel use under the hypotheses $\mathcal{H}_0$ and $\mathcal{H}_1$, respectively. Note that we may write the joint distributions $\mathbb{P}_0$ and $\mathbb{P}_1$ as a multiplication of the marginal distributions since both the channel uses and the code are $i.i.d.\ $ In particular, under $\mathcal{H}_0$ and given $p_j$, $f^0_{y_w[i]}$ is $\mathcal{CN}(0,\sigma_w^0)$, and under $\mathcal{H}_1$ and given $p_j$, $f^1_{y_w[i]}$ is $\mathcal{CN}(0,\sigma_w^1)$ where,
\begin{equation}\label{equ-channel distributions variances}
\begin{aligned}
\sigma_w^0&=\sigma_w^2+p_j\b{h}_{jw}^\dagger \bs{\Sigma} \b{h}_{jw},\\
\sigma_w^1&=\sigma_w^2+p_j\b{h}_{jw}^\dagger \bs{\Sigma} \b{h}_{jw}+P_a |h_{aw}|^2.
\end{aligned}
\end{equation}
The terms in the last line above reflect the self-noise power of Willie, the received AN power and the transmission power of Alice, respectively. It would be beneficial to express $\sigma_w^i, \ i\in\{0,1\}$ in \eqref{equ-channel distributions variances} using the representation $\bs{\Sigma}=\b{V} \b{X} \b{V}^\dagger$ as follows
\begin{equation}\label{equ-channel distributions variances SVD}
\begin{aligned}
\sigma_w^0&=\sigma_w^2+p_j\b{h}_{jw}^\dagger \b{V} \b{X} \b{V}^\dagger \b{h}_{jw},\\
\sigma_w^1&=\sigma_w^2+p_j\b{h}_{jw}^\dagger \b{V} \b{X} \b{V}^\dagger \b{h}_{jw}+P_a |h_{aw}|^2.
\end{aligned}
\end{equation} 

In \cite{sobers2017covert}, by using Fisher-Neyman factorization and likelihood ratio ordering techniques, it was shown that the above optimal ratio test is eventually an energy test on the average received power. This applies to our model as well, since from Willie's point of view, the models are the same. Specifically, the average received power, $P_w^{r_{av}}$, is compared with a threshold $\tau$,
\begin{equation}\label{equ-likelihood ratio test of Wille} 
P_w^{r_{av}} \triangleq \frac{1}{n} \sum_{i=1}^{n} |y_w[i]|^2 \Vectorstack{\mathcal{H}_1 > < \mathcal{H}_0}  \tau.
\end{equation}
One can realize that, given $p_j$, the average received power $P_w^{r_{av}}$ is a Gamma r.v. with parameters $k=n$ and $\theta=\frac{\sigma_w^i}{n}$ for $i=0,1$, i.e. $P_w^{r_{av}}\sim \Gamma(n,\frac{\sigma_w^i}{n})$. Furthermore, as $n\rightarrow \infty$, by the weak law of large numbers, $P_w^{r_{av}}$ converges in probability to $\sigma_w^i$.

\emph{Covertness achieving $P_a$:}
Willie compares $P_w^{r_{av}}$ to a threshold $\tau$; however, this threshold depends on the distribution of $P_j$ and thus may be optimized by Willie. The following analysis shows that for any threshold $\tau$ that Willie sets for himself, which is not known to anyone but Willie, there exists a construction by Alice such that \eqref{equ-covert criteria} holds. Specifically, we bound each of the probabilities $P_{MD}$ and $P_{FA}$ for a given value of $p_j$ resulting with sufficient conditions for covertness. Following similar conditioning arguments as in \cite{sobers2017covert}, for a fixed $p_j$, we have
\begin{equation}
P_{FA}(p_j)=\text{P}_\text{r}(P_w^{r_{av}} \geq \tau | \mathcal{H}_0, p_j).
\end{equation}
Recall that given $p_j$ and $\mathcal{H}_0$, $P_w^{r_{av}}\sim Gamma(n,\frac{\sigma_w^0}{n})$ and as $n\rightarrow \infty$, $P_w^{r_{av}}$ converges in probability to $\sigma_w^0$. Thus, for any given covertness requirement $\epsilon>0$, one can find $\delta^0(\epsilon,n)\rightarrow 0$ as $n\rightarrow \infty$ such that
\begin{equation*}
\text{P}_\text{r}(\sigma_w^0 - \delta^0 \leq P_w^{r_{av}} \leq \sigma_w^0+\delta^0| \mathcal{H}_0, p_j) > 1-\frac{\epsilon}{2}.
\end{equation*}
Note that the dependence of $\delta^0$ on $\epsilon$ and $n$ is omitted to ease notation. Since 
\begin{equation*}
\begin{aligned}
&\text{P}_\text{r}(P_w^{r_{av}} \geq \sigma_w^0-\delta| \mathcal{H}_0, p_j) \\
&\quad\quad\geq \text{P}_\text{r}(P_w^{r_{av}} \geq \sigma_w^0-\delta^0| \mathcal{H}_0, p_j) \\
&\quad\quad > \text{P}_\text{r}(\sigma_w^0 - \delta^0 \leq P_w^{r_{av}} \leq \sigma_w^0+\delta^0| \mathcal{H}_0, p_j),
\end{aligned}
\end{equation*}
for any $\delta(\epsilon) \geq \delta^0(\epsilon)$, then for any $\tau<\sigma_w^0-\delta(\epsilon)$ we have
\begin{equation}\label{equ-False alarm prob}
\text{P}_\text{r}(P_w^{r_{av}} \geq \tau | \mathcal{H}_0, p_j) > 1-\frac{\epsilon}{2}.
\end{equation}
Similarly for $P_{MD}$ given $p_j$,
\begin{equation}
P_{MD}(p_j)=\text{P}_\text{r}(P_w^{r_{av}} \leq \tau | \mathcal{H}_1, p_j).
\end{equation}
Now, $P_w^{r_{av}}\sim Gamma(n,\frac{\sigma_w^1}{n})$ and as $n\rightarrow \infty$, $P_w^{r_{av}}$ converges in probability to $\sigma_w^0$. Thus, one can find $\delta^1(\epsilon,n)\rightarrow 0$ as $n\rightarrow \infty$ such that
\begin{equation*}
\text{P}_\text{r}(\sigma_w^1 - \delta^1 \leq P_w^{r_{av}} \leq \sigma_w^1+\delta^1| \mathcal{H}_1, p_j) > 1-\frac{\epsilon}{2}.
\end{equation*}
Again, since 
\begin{equation*}
\begin{aligned}
&\text{P}_\text{r}(P_w^{r_{av}} \leq \sigma_w^1+\delta| \mathcal{H}_1, p_j) \\
&\quad\quad\geq \text{P}_\text{r}(P_w^{r_{av}} \leq \sigma_w^1+\delta^1| \mathcal{H}_1, p_j) \\
&\quad\quad > \text{P}_\text{r}(\sigma_w^0 - \delta^1 \leq P_w^{r_{av}} \leq \sigma_w^0+\delta^1| \mathcal{H}_1, p_j),
\end{aligned}
\end{equation*}
for any $\delta(\epsilon) \geq \delta^1(\epsilon)$, then for any $\tau>\sigma_w^1+\delta(\epsilon)$ we have
\begin{equation}\label{equ-miss detect prob}
\text{P}_\text{r}(P_w^{r_{av}} \leq \tau | \mathcal{H}_1, p_j) > 1-\frac{\epsilon}{2}.
\end{equation}
Set $\delta(\epsilon,n)=\max\{\delta^0(\epsilon,n),\delta^1(\epsilon,n)\}$ where we note that $\delta(\epsilon,n)\rightarrow 0$ as $n \rightarrow \infty$ as well. Since $\sigma_w^0,\sigma_w^1$ and $\delta$ implicitly depend on $p_j$, define the set of intervals $\mathcal{P}=\{p_j : \sigma_w^0-\delta<\tau<\sigma_w^1+\delta\}$. Thus, for all $p_j \notin \mathcal{P}$, either \eqref{equ-False alarm prob} or \ref{equ-miss detect prob} are satisfied and we have, 
\begin{equation}
P_{MD}(p_j)+P_{FA}(p_j)\geq 1-\frac{\epsilon}{2}.
\end{equation}
Since $P_j$ is a uniform r.v. we have 

\small
\begin{equation}\label{eq-probability of interval for p_j}
\begin{aligned}
\text{P}_\text{r}(\mathcal{P})&=\text{P}_\text{r}\left(\sigma_w^0-\delta<\tau<\sigma_w^1+\delta\right)\\
&=\text{P}_\text{r}\left( \frac{\tau-\sigma_w^2-P_a |h_{aw}|^2-\delta}{\b{h}_{jw}^\dagger \b{V} \b{X} \b{V}^\dagger \b{h}_{jw}} \leq P_j \leq   \frac{\tau-\sigma_w^2+\delta}{\b{h}_{jw}^\dagger \b{V} \b{X} \b{V}^\dagger \b{h}_{jw}} \right)\\
&\leq \frac{P_a |h_{aw}|^2+2\delta}{P_{max}\b{h}_{jw}^\dagger \b{V} \b{X} \b{V}^\dagger \b{h}_{jw}},
\end{aligned}
\end{equation}
\normalsize
where the inequality in the last line is due to the possibility that the boundries in the probability computation are outside the support of $P_j$ (i.e. outside $[0,P_{max}]$).
Note that the uniformly assumption on $P_j$ is critical for the result being independent of $\tau$.
Therefore, if we set $P_a=\frac{\epsilon P_{max}}{4|h_{aw}|^2} \b{h}_{jw}^\dagger \b{V} \b{X} \b{V}^\dagger \b{h}_{jw}$ and $\delta(\epsilon)=\frac{\epsilon P_{max}}{8}\b{h}_{jw}^\dagger \b{V} \b{X} \b{V}^\dagger \b{h}_{jw}$, for sufficiently large $n$, we are are left with 
\begin{equation*}
\text{P}_\text{r}(\mathcal{P})\leq\frac{\epsilon}{2}.
\end{equation*}
Considering all the above in order  we have,
\begin{equation}
\begin{aligned}
P_{MD}+P_{FA}&=\EX_{P_j}[P_{MD}(P_j)+P_{FA}(P_j)]\\
&\geq\EX_{P_j}[P_{MD}(P_j)+P_{FA}(P_j)|\mathcal{P}^c]\text{P}_\text{r}(\mathcal{P}^c)\\
&>1-\epsilon.
\end{aligned}
\end{equation}
The above shows that as long as Alice transmits with power $P_a=\frac{\epsilon P_{max}}{4|h_{aw}|^2} \b{h}_{jw}^\dagger \b{V} \b{X} \b{V}^\dagger \b{h}_{jw}$ the system is covert. The rate of Alice can be obtained by using $P_a$ above in Bob's SNR, which can be lower bounded by a constant, providing a positive rate.
\end{IEEEproof}

Lemma \ref{lem-Alice power for covert system} provides the transmission power of Alice as a function of Willie's CSI and the jammer's strategy, in a form of a multiplicative gain $\b{h}_{jw}^\dagger \b{V} \b{X} \b{V}^\dagger \b{h}_{jw}/|h_{aw}|^2$. This multiplicative gain is with respect to the case of a single antenna jammer as given in \cite{sobers2017covert}, which equals $P_a=\frac{P_{max}\epsilon}{4}$ if we adjust their model assumptions to ours. Note that although $h_{aw}$ and $\b{h}_{jw}$ are known, in the beginning of each slot, they are redrawn and thus $P_a$ should be updated accordingly. This means that the shared codebook needs to be updated as well. However, in the next section we provide analysis for this random gain and the corresponding received SNR. Using this analysis, one can first realize that the probability of this gain to be equal to zero, is zero. This eventually ensures a positive covert rate. Furthermore, one can compute the probability of this gain to be above some value. That is, define an outage criteria such that Alice would not transmit if the rate is too low. Using this criteria, Alice and Bob can agree on a covert rate.

Our main results for $M=1$ are presented in the following subsections with respect to the cases of known or unknown CSI of Willie.

\subsection{Known CSI of Willie}
\begin{theorem}\label{the-optimal solution for jammer covariance when helping Alice}
\emph{Assume a block fading AWGN channel and a jammer with $N$ antennas, who transmits AN with covariance matrix $\bs{\Sigma}$ and a total transmission power of $p_j$. For a given covertness requirement $\epsilon>0$ with globally known CSI of Willie, the optimal strategy for the jammer, i.e., the solution for the maximization problem in \eqref{equ-maximization problem of the jammer M=1 sub} while Alice is transmitting with the power given in Lemma \ref{lem-Alice power for covert system}, is the following power allocation,
\begin{equation}\label{equ-optimal power allocation sigma}
\bs{\Sigma}= \b{v}^*{\b{v}^*}^\dagger,
\end{equation}
where 
\begin{equation}\label{equ-optimal v direction full CSI}
\b{v}^*=\frac{\b{q}}{\|\b{q}\|}
\end{equation}
and $\b{q}$ is the eigenvector which corresponds to the highest eigenvalue of the matrix
\begin{equation}\label{equ-matrix for optimal q direction}
\left(\b{h}_{jb}  \b{h}_{jb}^\dagger +\sigma\b{I}\right)^{-1}\left(\b{h}_{jw} \b{h}_{jw}^\dagger \right), 
\end{equation}
where $\sigma=\frac{\sigma_b^2}{p_j}$.}
\end{theorem}

Hence the optimal strategy is to transmit AN in a single direction. The optimal direction, $\b{v}^*$, depends on both channel vectors $\b{h}_{jw}$ and $\b{h}_{jb}$. While it is hard to gain insight on the actual direction from \eqref{equ-matrix for optimal q direction} one can gain intuition from Equation \eqref{equ-intuition for direction} in the proof of Theorem \ref{the-optimal solution for jammer covariance when helping Alice}, where, it is clearly seen that $\b{v}^*$, on one hand, should be close to the direction of Willie, i.e., maximize the projection on $\b{h}_{jw}$, while on the other hand it should be orthogonal to Bob as much as possible, i.e., minimize the projection on $\b{h}_{jb}$. Figure \ref{fig-optimalDirectionPlot} depicts a visualization for a specific channel realization for $N=3$ antennas. One can easily observe that the optimal direction is indeed pointed towards Willie while being close to orthogonal to the direction of Bob.

\begin{IEEEproof}[Proof of Theorem \ref{the-optimal solution for jammer covariance when helping Alice}]
Setting Alice's transmission power in the maximization problem \eqref{equ-maximization problem of the jammer M=1 sub} according to \eqref{equ-Alice power for covert system} guarantees that the transmission is covert. Then, we have
\begin{equation}\label{equ-maximization problem of the jammer helping Alice}
\max_{\b{V,X}} \frac{\frac{\epsilon P_{max}}{4|h_{aw}|^2}\b{h}_{jw}^\dagger \b{V}\b{X} \b{V}^\dagger \b{h}_{jw}| h_{ab}|^2}{p_j\b{h}_{jb}^\dagger \b{V} \b{X} \b{V}^\dagger \b{h}_{jb}+\sigma_b^2}.
\end{equation}
To solve the above, we first find the optimal $\b{X}$ for any given $\b{V}$. Accordingly, for a fixed $\b{V}$, similar to \cite{shafiee2007achievable}, we further simplify \eqref{equ-maximization problem of the jammer helping Alice} as follows
\begin{equation}\label{equ-maximization problem of the jammer helping Alice simplified} 
=c_1\frac{P_{max}}{p_j}\max_{\b{X}} \frac{\b{w}^\dagger \b{X}\b{w}}{\b{b}^\dagger \b{X}\b{b}+\sigma},
\end{equation}
where $c_1=\epsilon| h_{ab}|^2/4|h_{aw}|^2$ depends on Alice's channels, $\sigma=\frac{\sigma_b^2}{p_j}$, $\b{w}=\b{V}^\dagger\b{h}_{jw}$ and $\b{b}=\b{V}^\dagger\b{h}_{jb}$.
Note that 
\begin{equation*}
\begin{aligned}
& \frac{\b{w}^\dagger \b{X}\b{w}}{\b{b}^\dagger \b{X}\b{b}+\sigma} = \frac{\sum_{l=1}^N \xi_l w_l^2 }{\sum_{l=1}^N \xi_l b_l^2 +\sigma}.
 \end{aligned}
\end{equation*}

 \begin{figure}[t]
   \centering
   	\includegraphics[width=0.35\textwidth]{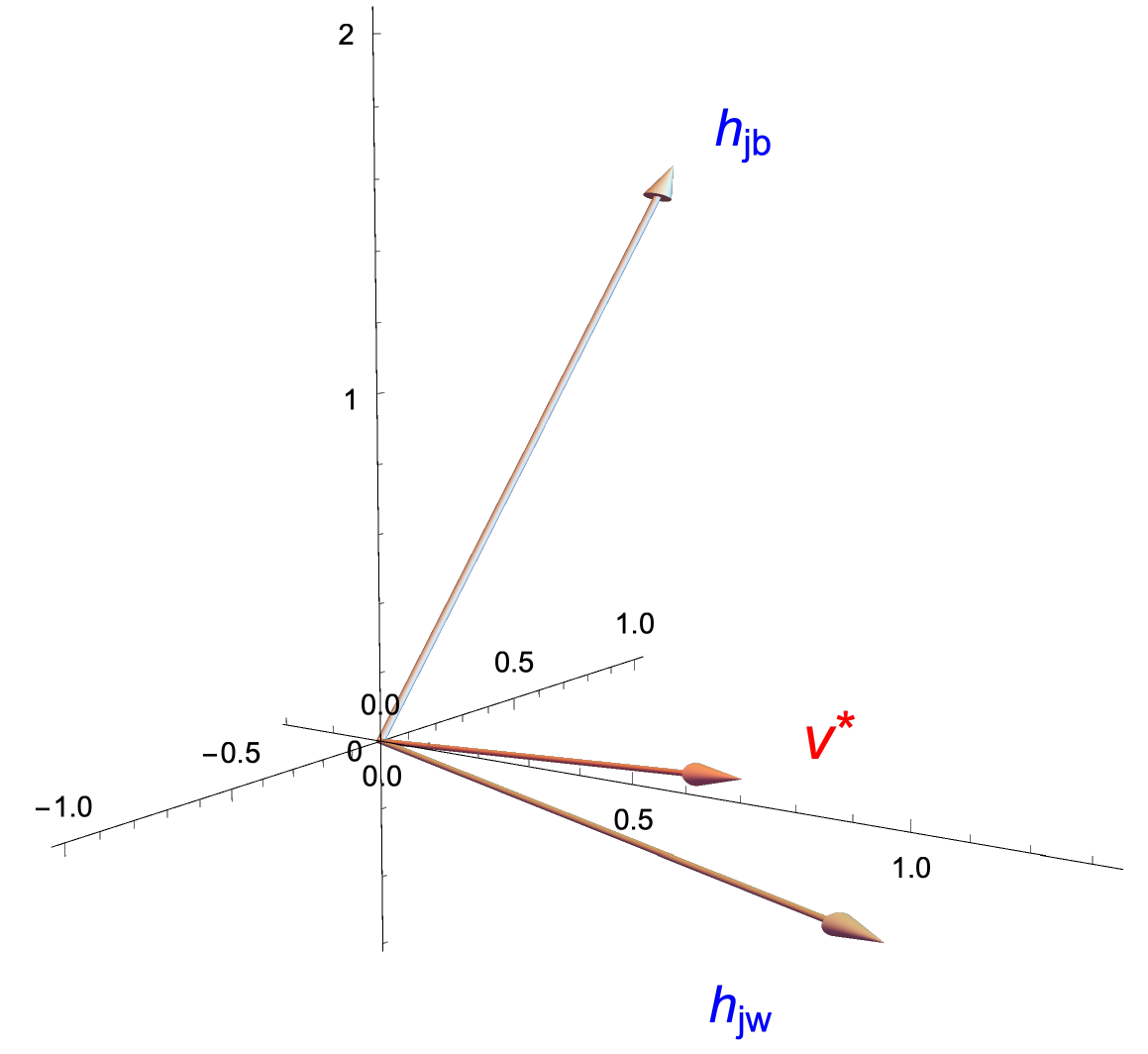}
   	\caption{An example for the optimal direction of the AN transmission by the jammer for a specific realization of the channel fading coefficients for the case of $M=1$ and $N=3$.}
	 \label{fig-optimalDirectionPlot}
 \end{figure}

We wish to show that the optimal power allocation $\bs{\xi}^*$ (which defines $\b{X}$), for a fixed $\b{w}$ and $\b{b}$, is a unit vector. To this end, we examine two indices $i$ and $j$ in $\bs{\xi}^*$ which have power allocations $(\xi_i,\xi_j)$ such that $\xi_i+\xi_j=\rho_{ij}$, where $\rho_{ij}>0$ is some constant. We will show first that either $\xi_i=\rho_{ij}$ or $\xi_j=\rho_{ij}$ must occur, hence, eventually, the optimal power allocation is a unit vector (since for any two indices $i$ and $j$, it is always better to move the allocation from one to the other, until one is exhausted).

The optimization problem on $\bs{\xi}$ can be written as follows,
\begin{equation*}
\begin{aligned}
 \max_{\bs{\xi}} f(\bs{\xi})
& \triangleq \max_{\bs{\xi}} \frac{\sum_{l=1}^N \xi_l w_l^2 }{\sum_{l=1}^N \xi_l b_l^2 +\sigma}\\
 &= \max_{\bs{\xi}} \frac{\sum_{l\neq i,j} \xi_l w_l^2 +w_i^2\xi_i+w_j^2\xi_j}{\sum_{l\neq i,j} \xi_l b_l^2 + b_i^2\xi_i+b_j^2\xi_j+\sigma}\\
 &= \max_{\bs{\xi}} \frac{\sum_{l\neq i,j} \xi_l w_l^2 +w_i^2\xi_i+w_j^2(\rho_{ij}-\xi_i)}{\sum_{l\neq i,j} \xi_l b_l^2 + b_i^2\xi_i+b_j^2(\rho_{ij}-\xi_i)+\sigma}\\
 &= \max_{\bs{\xi}} \frac{\sum_{l\neq i,j} \xi_l w_l^2 +\xi_i(w_i^2-w_j^2)+w_j^2\rho_{ij}}{\sum_{l\neq i,j} \xi_l b_l^2 + \xi_i(b_i^2-b_j^2)+b_j^2\rho_{ij}+\sigma}.\\
 \end{aligned}
\end{equation*}
Taking the derivative of the above according to $\xi_i$ shows that the function $f(\bs{\xi})$ is either monotonically increasing or monotonically decreasing with $\xi_i$ depending on the sign of 
$(w_i^2-w_j^2)\left(\sum_{l\neq i,j} \xi_l b_l^2+ \sigma +b_j^2\rho_{ij}\right) -(b_i^2-b_j^2)\left(\sum_{l\neq i,j} \xi_l w_l^2+w_j^2\rho_{ij}\right).$

Thus, for every two indices $i,j$, if $f(\bs{\xi})$ is monotonically decreasing in $\xi_i$, set $\xi_j=\rho_{ij}$. On the other hand, if $f(\bs{\xi})$ is monotonically increasing in $\xi_i$, set $\xi_j=0$. This essentially means that the optimal strategy of the jammer is to allocate all his power towards a single direction. 

In order to find this direction, which is the corresponding eigenvector $\b{v}$, we may write the unit rank matrix $\bs{\Sigma}$ as $\bs{\Sigma}=\b{v}\b{v}^\dagger$. Note that $\b{v}$ is constrained to have a unit norm, i.e., $\b{v}^\dagger\b{v}=1$. Returning to the maximization problem in \eqref{equ-maximization problem of the jammer helping Alice}, we have,
\begin{align}
& c_1\frac{P_{max}}{p_j}\max_{\b{V,X}} \frac{\b{h}_{jw}^\dagger \b{V}\b{X} \b{V}^\dagger \b{h}_{jw}}{\b{h}_{jb}^\dagger \b{V}\b{X} \b{V}^\dagger \b{h}_{jb}+\sigma}\\
& \quad\quad\quad =c_1\frac{P_{max}}{p_j}\max_{\b{v}} \frac{\b{h}_{jw}^\dagger \b{v}\b{v}^\dagger \b{h}_{jw}}{\b{h}_{jb}^\dagger \b{v}\b{v}^\dagger \b{h}_{jb}+\sigma\b{v}^\dagger\b{v}}\label{equ-optimizing v}\\
& \quad\quad\quad =c_1\frac{P_{max}}{p_j}\max_{\b{v}} \frac{\b{v}^\dagger\b{h}_{jw} \b{h}_{jw}^\dagger \b{v} }{\b{v}^\dagger \b{h}_{jb}  \b{h}_{jb}^\dagger \b{v}+\sigma\b{v}^\dagger\b{v}}\label{equ-intuition for direction}\\ 
& \quad\quad\quad =c_1\frac{P_{max}}{p_j}\max_{\b{v}} \frac{\b{v}^\dagger\b{h}_{jw} \b{h}_{jw}^\dagger \b{v} }{\b{v}^\dagger ( \b{h}_{jb}  \b{h}_{jb}^\dagger +\sigma\b{I})\b{v}}\\ 
&\quad\quad\quad  =c_1\frac{P_{max}}{p_j}\max_{\b{v}} \frac{\b{v}^\dagger \b{W} \b{v} }{\b{v}^\dagger \b{B} \b{v}},
\end{align}
where, $\b{W}=\b{h}_{jw} \b{h}_{jw}^\dagger, \quad\text{and,}\quad \b{B}= \b{h}_{jb}  \b{h}_{jb}^\dagger +\sigma\b{I}$. 
The above maximiztion problem is also known as the Rayleigh quotient \cite{mathai1992quadratic}, which is maximized by the eigenvector that corresponds to the highest eigenvalue of the matrix $\b{B}^{-1}\b{W}$. Denote this vector by $\b{q}$, the optimal $\b{v}$ is,
\begin{equation*}
\b{v}^*=\frac{\b{q}}{\|\b{q}\|}.
\end{equation*}
 \end{IEEEproof}

\subsection{Unknown CSI of Willie}\label{sec-unknown CSI Willie M=1}

We revisit Equation \eqref{equ-Alice power for covert system}. When $\b{h}_{jw}$ and $h_{aw}$ are unknown to Alice, she cannot determine her transmission power in order to stay covert. Moreover, without $\b{h}_{jw}$, it is not clear how to perform the optimization on $\b{h}_{jw}^\dagger \bs{\Sigma} \b{h}_{jw}$, in order to maximize Bob's SNR. We thus take the following course of action. We first lower bound Bob's SNR by a separable function, that will facilitate the optimization on $\bs{\Sigma}$. We then provide a sufficient condition for Alice's power, independent of $\b{h}_{jw}$, yet using the optimal $\bs{\Sigma}$ which resulted from the above optimization. The result is a strictly positive transmission power for Alice, a strictly positive SNR of Bob, $\epsilon-$covert transmission, with the optimal jamming strategy given the lower bound.


Accordingly, the following proposition provides a global lower bound on Bob's SNR, for any realization of the channel's coefficients, if Alice uses the same code construction as given in the proof of Lemma \ref{lem-Alice power for covert system}. 
\begin{proposition}\label{pro-lower bound on Bob SNR}
\textit{Using the code construction given in the proof of Lemma \ref{lem-Alice power for covert system}, for any realization of Willie's CSI, Bob's received SNR is lower bounded by
\begin{equation}\label{equ-lower bound on Bob SNR}
 \frac{c_1P_{max}}{p_j} \frac{\|\b{h}_{jw}\|^2 \min\limits_{i=1,...,d}\xi_i }{\|\b{h}_{jb}\|^2 \max\limits_{i=1,...,d}\xi_i+\sigma}. 
\end{equation}}
\end{proposition}
\begin{IEEEproof}
Setting Alice's transmission power in the maximization problem \eqref{equ-maximization problem of the jammer M=1 sub} according to \eqref{equ-Alice power for covert system} while using the notations of Theorem \ref{the-optimal solution for jammer covariance when helping Alice}, we have
\begin{equation}
\frac{c_1P_{max}}{p_j} \frac{\b{h}_{jw}^\dagger \b{V}\b{X} \b{V}^\dagger \b{h}_{jw}}{\b{h}_{jb}^\dagger \b{V} \b{X} \b{V}^\dagger \b{h}_{jb}+\sigma}.
\end{equation}
For any vector $\b{h}$ and matrices $\b{V}, \b{X}$, we have \cite[Theorem 2.4.3]{mathai1992quadratic}:
\begin{equation}\label{eq-upper and lower bound on quadratic form}
\|\b{h}\|^2 \min_{i=1,...,d}\xi_i \leq \b{h}^\dagger \b{V} \b{X} \b{V}^\dagger \b{h}  \leq \|\b{h}\|^2  \max_{i=1,...,d}\xi_i,
\end{equation}
and the bound follows immediately.
\end{IEEEproof}


For covert communication, a natural choice would be to maximize the lower bound, for example, to ensure that the SNR is above a certain level with very high probability. This may be a very important promise since it enables Alice to communicate while remaining undetected. We thus assume for the rest of this section that the jammer strategy is subject to maximizing the lower bound on Bob's SNR given in \eqref{equ-lower bound on Bob SNR}. The following proposition provides the jammer strategy for this case.
\begin{proposition}\label{pro-d direction with equal power}
\textit{To maximize the lower bound in \eqref{equ-lower bound on Bob SNR}, the jammer should direct the AN towards $d$ directions, with all of his available power equally divided between them. That is  $\bs{\Sigma}=\b{V}\b{X}\b{V}^\dagger$, with $\b{V}$ being a unitary matrix of dimension $N\times d$ with rank $d$ and $\b{X} = \frac{1}{d} \b{I}_{d\times d}$.}
\end{proposition}
\begin{IEEEproof}
The proof follows since $\sum_i^d \xi_i=1$ and we wish to maximize the smallest $\xi_i$ and minimize the highest $\xi_i$, thus letting $\xi_i=\frac{1}{d}$ for all $i$ will maximize the lower bound in \eqref{equ-lower bound on Bob SNR}. 
\end{IEEEproof}


Note that the above proposition is consistent with Theorem \ref{the-optimal solution for jammer covariance when helping Alice}, in which the jammer transmits in a single direction, i.e., $d=1$ for the case of known CSI. On the other hand, when the CSI of Willie is unknown ($\b{h}_{jw}$ and $h_{aw}$ are not known and random), to maximize the SNR, the jammer strategy should follow Proposition \ref{pro-d direction with equal power}. In fact, transmitting AN uniformly in all directions is consistent with AN transmission strategies for the case of unknown CSI towards the adversary in the context of wiretap channel, e.g., \cite{khisti2007secure,goel2008guaranteeing}. However, while the reason for transmitting uniformly across the basis of the null-space of Bob was intuitive in \cite{khisti2007secure,goel2008guaranteeing}, herein we prove that this choice actually maximises the lower bound on Bob's SNR.

Considering the above, and similarly to the previous subsection, we first express Alice's transmission power, $P_a^c(\b{V},\b{X})$, which assures covertness for the case of unknown CSI of Willie. 

\begin{lemma}\label{lem-Alice power for covert system with no CSI}
\emph{Assume a block fading AWGN channel and a jammer with $N$ antennas, who transmits AN with his total transmission power, $p_j$, equally divided into $d$ arbitrary directions. I.e., $rank(\b{X})=d$. For a given covertness requirement $\epsilon >0$ when the CSI of Willie is unknown, as long as Alice transmits with power
\begin{equation}\label{equ-Alice power for covert system with no CSI}
P_a\leq \frac{\epsilon P_{max}}{12 e \ln{\frac{6}{\epsilon}}}\left(\frac{\epsilon}{6}\right)^{\frac{1}{d}},
\end{equation}
the system is covert. I.e., \eqref{equ-covert criteria} applies and Alice's rate is strictly positive.}
\end{lemma}
\begin{IEEEproof}[Proof of Lemma \ref{lem-Alice power for covert system with no CSI}]
The construction is similar to the one given in Lemma \ref{lem-Alice power for covert system}. However, now, to obtain the covertness achieving power $P_a$, one needs to consider the channel's statistics for $h_{aw}$ and $\b{h}_{jw}$ as well as $P_j$ when computing the probability $\text{P}_\text{r}(\mathcal{P})$ in equation \eqref{eq-probability of interval for p_j}. To do so, we upper bound $\text{P}_\text{r}(\mathcal{P})$ and find $P_a$ such that \eqref{equ-covert criteria} applies. Specifically, we can write for all $\alpha>0$
\begin{equation}
\begin{aligned}
\text{P}_\text{r}(\mathcal{P})&=\text{P}_\text{r}\left(\mathcal{P}\Big||h_{aw}|^2 > \alpha\right)\text{P}_\text{r}\left(|h_{aw}|^2 > \alpha\right)+\\
&\quad\quad\quad\quad\quad \text{P}_\text{r}\left(\mathcal{P}\Big||h_{aw}|^2\leq \alpha\right)\text{P}_\text{r}\left(|h_{aw}|^2\leq \alpha\right) \\
&\leq \text{P}_\text{r}\left(|h_{aw}|^2 > \alpha\right)+ \text{P}_\text{r}\left(\mathcal{P}\Big||h_{aw}|^2\leq \alpha\right).
\end{aligned}
\end{equation}
Denoting $\Psi=\b{h}_{jw}^\dagger \b{V} \b{X} \b{V}^\dagger \b{h}_{jw}$ we can further upper bound the above for all $\beta > 0$
\begin{equation}
\begin{aligned}
\text{P}_\text{r}\left(\mathcal{P}\right)&\leq \text{P}_\text{r}\left(|h_{aw}|^2 > \alpha\right)+ \text{P}_\text{r}\left(\mathcal{P}\Big||h_{aw}|^2\leq \alpha\right)\\
&= \text{P}_\text{r}\left(|h_{aw}|^2 > \alpha\right) \\
&\quad\quad +\text{P}_\text{r}\left(\mathcal{P}\Big||h_{aw}|^2\leq \alpha, \Psi < \beta\right)\text{P}_\text{r}\left(\Psi < \beta\right) \\
&\quad\quad + \text{P}_\text{r}\left(\mathcal{P}\Big||h_{aw}|^2\leq \alpha,\Psi \geq \beta\right)\text{P}_\text{r}\left(\Psi \geq \beta\right)\\
&\leq \text{P}_\text{r}\left(|h_{aw}|^2 > \alpha\right) + \text{P}_\text{r}\left(\Psi < \beta\right) \\
& \quad\quad\quad + \text{P}_\text{r}\left(\mathcal{P}\Big||h_{aw}|^2\leq \alpha,\Psi \geq \beta\right).
\end{aligned}
\end{equation}
Recall the definition of $\mathcal{P}=\{p_j : \sigma_w^0-\delta<\tau<\sigma_w^1+\delta\}$. If we substitute $\alpha$ and $\beta$ for $|h_{aw}|^2$ and $\Psi$, respectively, we can upper bound the third term in the equation above and thus increase the probability. That is, set $\alpha$ and $\beta$ in \eqref{eq-probability of interval for p_j}. This replacement can only increase the length of the considered interval. Since the distribution is uniform, we have
\begin{equation}
\begin{aligned}
\text{P}_\text{r}\left(\mathcal{P}\right)&\leq  \text{P}_\text{r}\left(|h_{aw}|^2 > \alpha\right) + \text{P}_\text{r}\left(\Psi < \beta\right) \\
& \quad\quad\quad + \text{P}_\text{r}\left(\mathcal{P}\Big||h_{aw}|^2\leq \alpha,\Psi \geq \beta\right)\\
& \leq  \text{P}_\text{r}\left(|h_{aw}|^2 > \alpha\right) + \text{P}_\text{r}\left(\Psi < \beta\right) \\
&\quad +\text{P}_\text{r}\left( \frac{\tau-\sigma_w^2-P_a \alpha-\delta}{\beta} \leq P_j \leq   \frac{\tau-\sigma_w^2+\delta}{\beta} \right)\\
& \leq \text{P}_\text{r}\left(|h_{aw}|^2 > \alpha\right) + \text{P}_\text{r}\left(\Psi < \beta\right) + \frac{2\delta+P_a\alpha}{\beta P_{max}}.
\end{aligned}
\end{equation}
Since $|h_{aw}|^2$ is a $\chi^2(2)$ r.v., by setting $\alpha=2\ln\frac{6}{\epsilon}$ we have, 
\begin{equation}
\text{P}_\text{r}\left(|h_{aw}|^2 > \alpha\right) = \frac{\epsilon}{6}.
\end{equation}
To obtain the distribution of $\Psi$, recall that the elements of the diagonal matrix $\b{X}$ are all equal to $1/d$, hence, $\Psi=\frac{1}{d}\b{h}_{jw}^\dagger \b{V} \b{I} \b{V}^\dagger \b{h}_{jw}=\frac{1}{d}\b{h}_{jw}^\dagger \b{V} \b{V}^\dagger \b{h}_{jw}$. Since $\b{V}$ is a unitary matrix, $\b{h}_{jw}^\dagger \b{V}$ is distributed as a complex Gaussian vector of length $d$ with independent variables.  Accordingly, $d\cdot\Psi$ is a $\chi^2_{(2d)}$ r.v.. This is also consistent with \cite[Theorem 2, Ch.1]{pavur1980quadratic}, using a different approach. Thus, we compute
\begin{equation}
\begin{aligned}
\text{P}_\text{r}\left(\Psi < \beta\right) &=\text{P}_\text{r}\left(d\Psi < d\beta\right)\\
 &\overset{(a)}{\leq} \left( \frac{d\beta}{2d} e^{\left(1-\frac{d\beta}{2d}\right)} \right)^d\\
&\overset{(b)}{\leq}  \left(  \frac{\beta}{2} e\right)^d\\
&\overset{(c)}{\leq}  \frac{\epsilon}{6}.\\
\end{aligned}
\end{equation}
Where $(a)$ follows from the Chernoff bound which requires that $0<\beta<2$ and $(b)$ follows for such $\beta$. $(c)$ is by setting $\beta=\frac{2}{e}\sqrt[d]{\frac{\epsilon}{6}}$. Note that this choice of $\beta$ satisfies the Chernoff's bound requirement for all $\epsilon\in(0,1)$ and $d\in\Z^+ \backslash \{0\}$.

Thus, setting $P_a=\frac{\beta}{\alpha}\frac{\epsilon}{12}P_{max}$ and $\delta=\frac{\epsilon}{24}\beta P_{max}$ we have,
\begin{equation*}
\text{P}_\text{r}\left(\mathcal{P}\right) \leq \frac{\epsilon}{2}.
\end{equation*}
Continuing similar to the proof of Lemma \ref{lem-Alice power for covert system}, the covertness criteria $P_{MD}+P_{FA}\geq 1-\epsilon$ holds. 
\end{IEEEproof}

In the following corollary, we present the covertness achieving transmission power for Alice in the case where the jammer has a single antenna (i.e., $N=1$). We note that this result is different from the one obtained in \cite{sobers2017covert}, as the model assumptions are different. In \cite{sobers2017covert}, it was assumed that Willie does not know his own CSI, thus the jammer may use AN with constant variance $P_j=P_{max}$. Of course, this eventually affects Alice's power. In this work, we assume a stringent model where Willie knows his own CSI, therefore the communicating parties cannot assume that the channel randomization is enough to confuse Willie. Thus, the jammer must employ AN with varying power.

\begin{corollary} 
\emph{Assume a block fading AWGN channel and a jammer with a single antenna and a total transmission power of $p_j$, where the CSI of Willie is unknown. For a given covertness requirement $\epsilon >0$, as long as Alice transmits with power
\begin{equation}
P_a\leq \frac{\ln{1-\frac{\epsilon}{6}}}{\ln{\frac{\epsilon}{6}}}\frac{\epsilon}{12}P_{max},
\end{equation}
the system is covert. I.e., \eqref{equ-covert criteria} applies and Alice's rate is strictly positive.}
\end{corollary}

\begin{IEEEproof}
Following the same steps as the proof of Lemma \ref{lem-Alice power for covert system with no CSI}, when the jammer is equipped with a single antenna, the received power of the AN at Willie is $p_j|h_{jw}|^2$. Thus, in the proof of Lemma \ref{lem-Alice power for covert system with no CSI}, $\Psi=|h_{jw}|^2$, which is distributed as a $\chi^2(2)$ r.v.. Therefore, we may set $\beta=2\ln{\frac{6}{6-\epsilon}}$, which results in $\text{P}_\text{r}\left(\Psi < \beta\right)=\frac{\epsilon}{6}$.
\end{IEEEproof}
A more strict upper bound for Alice's transmission power can be obtained following to the inequality $\frac{\ln{1-x}}{\ln{x}}>x\sqrt{x}$, for $0<x<1$, which results with $P_a\leq\left(\frac{\epsilon}{6}\right)^{2.5}\frac{P_{max}}{2}$.

\begin{corollary}\label{cor-p_a as a function of the jammer's strategy only}
\textit{Given the jammer's strategy $\bs{\Sigma}$, for the case of known and unknown CSI of Willie, for any fixed $\epsilon$ and $P_{max}$, a covertness achieving, strictly positive, power of Alice is attainable.} 
\end{corollary}
The proof follows immediately from the proofs of Lemmas \ref{lem-Alice power for covert system} and \ref{lem-Alice power for covert system with no CSI} for the case of known and unknown CSI of Willie, respectively. 

As Lemma \ref{lem-Alice power for covert system with no CSI} suggests, the covertness-achieving power of Alice for the case of unknown CSI, assuming equal power allocation, only depends on the number of directions the jammer transmits into. Again, such an assumption is motivated by the desire to promise minimal value for Alice's transmission power. Accordingly, the optimal strategy under such assumptions is given in the following theorem.

\begin{theorem}\label{the-optimal solution for jammer covariance when helping Alice no CSI}
\emph{Assume a block fading AWGN channel, and a jammer with $N$ antennas, who transmits AN with covariance matrix $\bs{\Sigma}$ such that his total transmission power, $p_j$, is equally divided into $d$ arbitrary directions. For a given covertness requirement $\epsilon>0$ with unknown CSI of Willie, the optimal strategy for the jammer, is the following power allocation (for $N\geq 2$ antennas),
\begin{equation}\label{equ-optimal power allocation sigma no CSI}
\bs{\Sigma}= \b{V}^*\b{X}^*{\b{V}^*}^\dagger,
\end{equation}
where in case $\sigma=\frac{\sigma_b^2}{p_j}<\frac{\|\b{h}_{jb}\|^2}{N\left( {\frac{\epsilon}{6}}^{-\frac{1}{N(N-1)}}-1 \right)}$,
\begin{equation}\label{equ-optimal V directions N-1}
\begin{aligned}
&\b{V}^*=\b{Q}\ \text{  and } 
&\b{X}^*=\frac{1}{N-1}I_{(N-1)\times(N-1)},
\end{aligned}
\end{equation}
otherwise,
\begin{equation}\label{equ-optimal V directions N}
\begin{aligned}
&\b{V}^*=  \left( \b{Q}\ ,\ \frac{\b{h}_{jb}}{\|\b{h}_{jb}\|} \right)\text{  and, } 
&\b{X}^*=\frac{1}{N}I_{N\times N},
\end{aligned}
\end{equation}
where $\b{Q}$ is a matrix whose columns span the null-space of $\b{h}_{jb}$.} 
\end{theorem}   

The above result implies that when Willie's CSI is unknown, the jammer should transmit his AN either to the null-space of Bob or isotropically to all directions. The choice depends on the magnitude of the channel $\b{h}_{jb}$. That is, for fixed $N,\sigma_b^2$, and $p_j$, if  $\|\b{h}_{jb}\|$ is high, the optimal strategy would be to transmit only to the null-space. On the other hand, when $\|\b{h}_{jb}\|$ is low, transmitting to all directions (including Bob's) would be better in the sense of maximizing the covert rate. An interpretation for the above is possible by considering the parameter $\sigma$ which reflects the ratio between the "bad" noise power that Bob endures due to his own antenna noise and the "good" AN power which helps Alice and consequently helps Bob's SNR. When $\sigma$ is low Bob endures more "bad" noise and thus the jammer should avoid adding more noise power in the direction of Bob.
Note also that Theorem \ref{the-optimal solution for jammer covariance when helping Alice no CSI} refers only to the case of a multiple antenna jammer (i.e. $N \geq 2$). $N=1$ is not considered since no optimization can be performed. Moreover, not transmitting AN is not possible as it results with zero covert rate (a square root law). 

\begin{figure}[t]
   \centering
   	\includegraphics[width=0.45\textwidth]{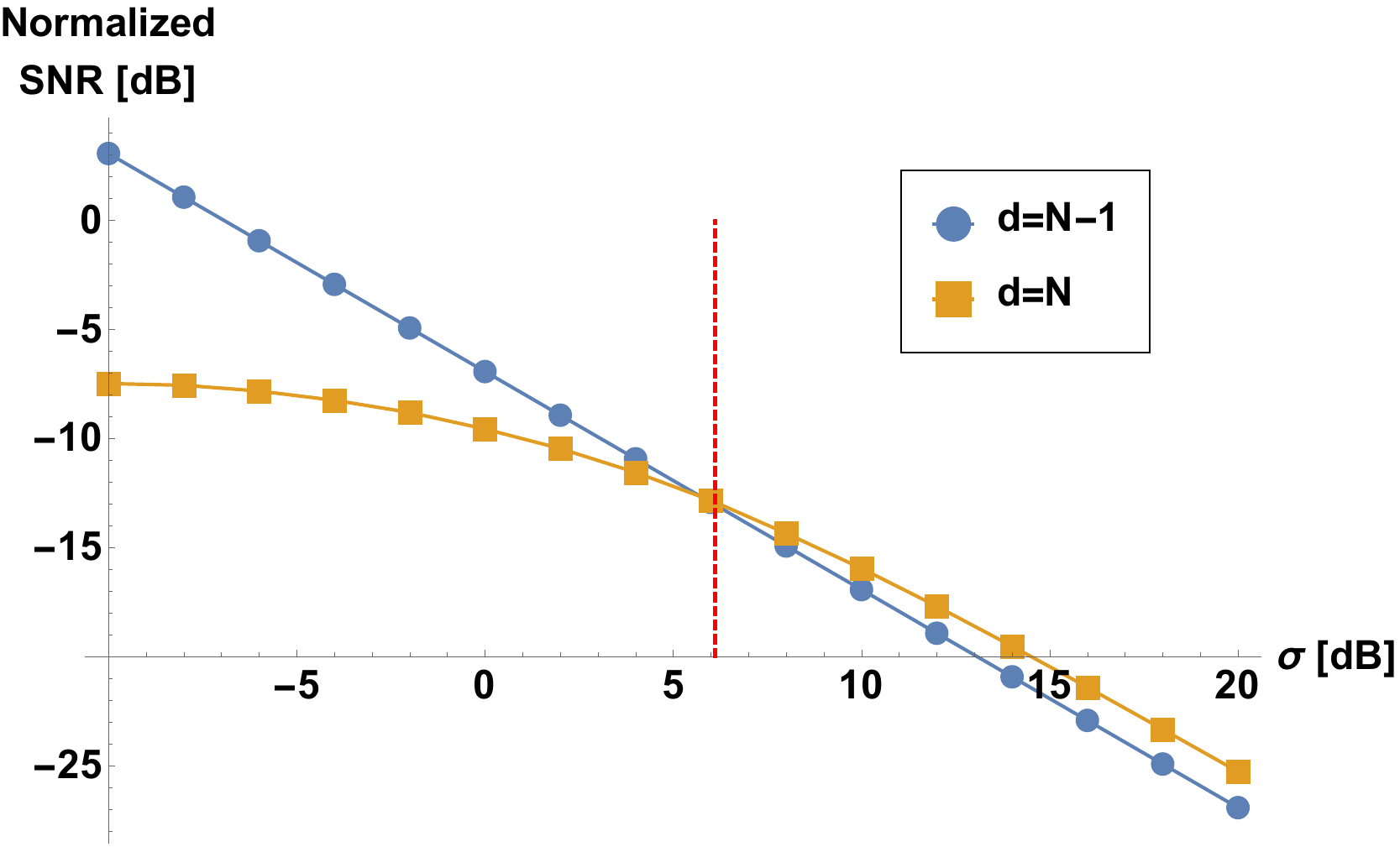}
   	\caption{Simulation results for Bob's received SNR as a function of $\sigma$ for $N=4$ antennas at the jammer, where the CSI of Willie is unknown. The SNR values are normalized so the leading constant coefficients is equal to 1.} 
	 \label{fig-SNR_null_full_space_M=1_NoCSI}
 \end{figure}

Figure \ref{fig-SNR_null_full_space_M=1_NoCSI} depicts the turning point of the average SNR value when comparing the strategy of transmitting the AN isotropically to all directions against transmitting only in the null-space. The simulation was performed for $\epsilon=0.05$ and $N=4$ antennas at the jammer.  As described in Theorem \ref{the-optimal solution for jammer covariance when helping Alice no CSI}, the figure depicts that around the point of $\sigma=\EX\left[\frac{\|\b{h}_{jb}\|^2}{N\left( {\frac{\epsilon}{6}}^{-1/N(N-1)}-1 \right)}\right]=\frac{8}{4\left( {\frac{\epsilon}{6}}^{-1/12}-1 \right)} \approx 6.1[dB]$\footnote{$\|\b{h}_{jb}\|^2$ is distributed as a $\chi^2_{2N}$ r.v.} the jammer should switch strategy in order to maximize the SNR. 

\begin{IEEEproof}[Proof of Theorem \ref{the-optimal solution for jammer covariance when helping Alice no CSI}]
Setting Alice's transmission power in the maximization problem \eqref{equ-maximization problem of the jammer M=1 sub} according to \eqref{equ-Alice power for covert system with no CSI} guarantees that the transmission is covert. Then, considering that $\b{X}$ is a diagonal matrix with $\xi_l = \frac{1}{d}$ for all $l$, we have
\begin{equation*}
\begin{aligned}
&\max_{\b{V},d} \frac{ \frac{\epsilon P_{max}}{12 e \ln{\frac{6}{\epsilon}}}\left(\frac{\epsilon}{6}\right)^{\frac{1}{d}} | h_{ab}|^2}{p_j\b{h}_{jb}^\dagger \b{V} \b{X} \b{V}^\dagger \b{h}_{jb}+\sigma_b^2},\\
& s.t. \quad \xi_l = \frac{1}{d}, \ \ l=1,...,d.
\end{aligned}
\end{equation*}
The above can be simplified to
\begin{equation}\label{equ-simplified opt prob no csi M=1}
c_2 \frac{P_{max}}{p_j} \max_{d,\b{b}} \frac{ \left(\epsilon' \right)^{\frac{1}{d}} }{\sum_{l=1}^d \frac{1}{d} b_l^2+\sigma},
\end{equation}
where $c_2=\frac{\epsilon' | h_{ab}|^2}{2 e \ln{\epsilon'}}$, $\epsilon'=\frac{\epsilon}{6}$, $\sigma=\frac{\sigma_b^2}{p_j}$ and $\b{b}= \b{V}^\dagger \b{h}_{jb}$. Let $\b{Q}$ be a matrix whose columns span the null-space of $\b{h}_{jb}$, and let $\b{V}$ be a matrix with any $d$ columns taken from the columns of $\b{Q}$, i.e., $\b{V}$ is of dimension $N\times d$. Since the numerator is an increasing function of $d$ and in the denominator we can have $b_l^2=0$ for $l=1,...,N-1$ due to the choice of $\b{V}$, any $d<N-1$, will result in a lower SNR. Thus, one needs to consider only the cases of $d=N-1$ and $d=N$, for which the target function in \eqref{equ-simplified opt prob no csi M=1} is equal to
\begin{equation}\label{equ-SNR values}
\begin{dcases*}
         \frac{ \left(\epsilon' \right)^{\frac{1}{N-1}} }{\sigma} & if $\ d=N-1$ \\
        \frac{ \left(\epsilon' \right)^{\frac{1}{N}} }{ \frac{1}{N} \|\b{h}_{jb}\|^2+\sigma} & if $\ d=N$. 
        \end{dcases*}
\end{equation}
When $\sigma>\frac{\|\b{h}_{jb}\|^2}{N\left( \epsilon'^{-1/N(N-1)}-1 \right)}$ the SNR is larger for the case where $d=N$. Thus, in this case, $\b{V}$ must span the whole space. So, by setting $\b{V}^*=\left( \b{Q} \ , \ \frac{\b{h}_{jb}}{\|\b{h}_{jb}\|} \right)$ we have $b_N=\|\b{h}_{jb}\|$ and the resulting value for the target function in \eqref{equ-SNR values}. Otherwise, it is better to transmit in $d=N-1$ directions and only in the null-space of Bob, i.e., $\b{V}^*=\b{Q}$ which completes the proof.
\end{IEEEproof}


\section{AN Transmission and detection Strategies for $M>1$ Antennas at Bob}\label{sec-AN Transmission Strategies M>1}

In this section, we analyze the AN transmission strategy of the jammer and the detection strategy of Bob in case Bob has $M>1$ antennas. When Bob is equipped with multiple antennas, the AN transmission and the detection strategies of the jammer and Bob are coupled and depend on the joint information both have.

Accordingly, we formulate a global optimization problem on both the transmission of the jammer and the detection of Bob where, similar to Section \ref{sec-AN Transmission Strategies M=1}, we restrict ourselves to constructions which satisfy the covertness requirement \eqref{equ-covert criteria} for the cases of known and unknown CSI of Willie, respectively.

Recall that Bob employs a linear receiver. Thus, following the SNR expression in \eqref{equ-The received SNR at Bob for a general projection c}, and assuming that the system is covert (i.e. Alice is transmitting with a covertness achieving power $P_a^c(\bs{\Sigma})$), we may write the global optimization problem as follows, 
\begin{equation}\label{equ-maximization problem of the jammer for general projection}
\begin{aligned}
&\max_{\b{V,X,c}} \ \frac{P_a^c(\b{V},\b{X})|\b{c}^\dagger\b{h}_{ab}|^2}{\b{c}^\dagger\left(p_j\b{H}_{jb} \b{V} \b{X} \b{V}^\dagger\b{H}_{jb}^\dagger  +\sigma_b^2\b{I} \right)\b{c}}\\
& s.t. \quad 0\leq\xi_l \leq 1 \text{ and,  } \sum_{l=1}^N \xi_l =1.
\end{aligned}
\end{equation}

Note that $\b{c}$ should be optimized together with $\bs{\Sigma}$ (i.e., with $\b{V}$ and $\b{X}$) which leads to yet another non-convex optimization problem. In the following subsections, we suggest strategies for Bob and the jammer which decentralized the global optimization above to obtain practical solutions for the cases of known and unknown CSI of Willie, respectively. The numerical results of the suggested strategies are shown to be very close to the optimal solution of \eqref{equ-maximization problem of the jammer for general projection}, which can be acquired numerically.

\subsection{Known CSI of Willie}

The following lemma asserts that regardless of Bob's filter, under the covertness construction given in Lemma \ref{lem-Alice power for covert system}, the jammer should direct his AN to a single direction.

\begin{lemma}\label{lem-optimal power allocation is a single beam in general}
\emph{For any linear filter $\b{c}$, the optimal strategy for the jammer, i.e., the solution to \eqref{equ-maximization problem of the jammer for general projection} when the CSI of Willie is known, is to transmit all his available power towards a single direction.}
\end{lemma}
\begin{IEEEproof}
Substituting Alice's transmission power given in \eqref{equ-Alice power for covert system} in \eqref{equ-maximization problem of the jammer for general projection}, the maximization problem in \eqref{equ-maximization problem of the jammer for general projection} can be reduced as follows,
\begin{equation}
\begin{aligned}
&\max_{\b{V,X,c}} |\b{c}^\dagger\b{h}_{ab}|^2\frac{\frac{\epsilon P_{max}}{4|h_{aw}|^2}\b{h}_{jw}^\dagger \b{V} \b{X} \b{V}^\dagger \b{h}_{jw}}{\b{c}^\dagger\left(p_j\b{H}_{jb} \b{V} \b{X} \b{V}^\dagger\b{H}_{jb}^\dagger  +\sigma_b^2\b{I} \right)\b{c}}\\
&=\max_{\b{V,X,c}} |\b{c}^\dagger\b{h}_{ab}|^2\frac{\frac{\epsilon P_{max}}{4|h_{aw}|^2}\b{h}_{jw}^\dagger \b{V} \b{X} \b{V}^\dagger \b{h}_{jw}}{\b{c}^\dagger p_j\b{H}_{jb} \b{V} \b{X} \b{V}^\dagger\b{H}_{jb}^\dagger \b{c} +\sigma_b^2\|\b{c}\|^2 }\\
&=\max_{\b{b}',\b{X},\b{c}} c_3 \frac{P_{max}}{p_j} |\b{c}^\dagger\b{h}_{ab}|^2\frac{\b{w}^\dagger \b{X}\b{w}}{ \b{b'}^\dagger \b{X} \b{b'} +\sigma\|\b{c}\|^2 },\\
\end{aligned}
\end{equation}
where $c_3=\epsilon/4|h_{aw}|^2$, $\sigma=\frac{\sigma_b^2}{P_j}$, $\b{w}=\b{V}^\dagger\b{h}_{jw}$ and $\b{b}'$ is of the form $\b{b}'=\b{V}^\dagger\b{H}_{jb}^\dagger \b{c}$ where $\b{V}$ is any unitary matrix. Following the same analysis performed in the proof of Theorem \ref{the-optimal solution for jammer covariance when helping Alice}, it can be shown that for any fixed $\b{w},\b{b'}$ and $\b{c}$ the optimal power allocation $\bs{\xi}^*$ (i.e., the diagonal of $\b{X}$) is a unit vector.
\end{IEEEproof}

Lemma \ref{lem-optimal power allocation is a single beam in general} agrees with the conclusion of Theorem \ref{the-optimal solution for jammer covariance when helping Alice} for a single direction strategy (when $M=1$) and shows that since Bob is equipped with several antennas yet a fixed filter, the jammer should transmit in a single direction. Accordingly, using Lemma \ref{lem-optimal power allocation is a single beam in general}, the maximization problem in \eqref{equ-maximization problem of the jammer for general projection} can be reduced as follows,

\begin{equation}\label{equ-maximization problem of the jammer for general projection reduced to single direction}
\begin{aligned}
&c_3 \frac{P_{max}}{p_j}\max_{\b{v,c}} \frac{\left(\b{h}_{jw}^\dagger \b{v} \b{v}^\dagger \b{h}_{jw}\right)|\b{c}^\dagger\b{h}_{ab}|^2}{\b{c}^\dagger\left(\b{H}_{jb}  \b{v} \b{v}^\dagger\b{H}_{jb}^\dagger  +\sigma\b{I} \right)\b{c}}.
\end{aligned}
\end{equation}
Inspired by existing works on MIMO communication, we suggest and analyze sub-optimal solutions for the above optimization problem. Specifically, we consider the null steering (also known as "masked beamforming") and Maximal Ratio Combiner (MRC) techniques (\cite{godara1997application,goldsmith2005wireless}) which Bob and the jammer may perform independently to simplify the optimization problem. That is, instead of solving the global optimization for both strategies simultaneously, we fix one and perform optimization on the other. In a way, these sub-optimal schemes decentralized the global optimization problem by forcing the jammer and Bob to rely only on themselves and their ability to enhance the covert rate. Later on, in Corollary \ref{cor-optimal solution for large sigma}, we show that some of the suggested strategies are asymptotically optimal as $\sigma$ grows.


In what follows we present analysis for the sub-optimal strategies described above. We solve the optimization problem in \eqref{equ-maximization problem of the jammer for general projection reduced to single direction} first for a fixed $\b{c}$ and then for a fixed $\b{v}$. Then, we suggest strategies for the jammer and Bob given a fixed $\b{c}$ or a fixed $\b{v}$. Specifically, we first consider a fixed linear filter, $\b{c}$, as an MRC filter, to match the channel between Alice and Bob, ignoring the AN. Then fix $\b{v}$ by the jammer, to be in the direction of Willie. Lastly, we consider the strategy of canceling the AN at Bob's receiver by either the jammer or Bob where, depending on the number of antennas each has, they may agree in advance upon the responsibility for the cancelation task.

\underline{\textbf{Fixed $\b{c}$:}} \newline
For any fixed $\b{c}$ we have the following theorem.
\begin{theorem}\label{the-optimal strategy for jammer for fixed c}
\emph{For a fixed linear filter $\b{c}$, the strategy for the jammer, i.e., the solution to \eqref{equ-maximization problem of the jammer for general projection reduced to single direction} when the CSI of Willie is known, is the following power allocation,
\begin{equation}
\bs{\Sigma}=\b{v}^*{\b{v}^*}^\dagger,
\end{equation}
where $\b{v}^*$ is the eigenvector which corresponds to the highest eigenvalue of the matrix $\b{B}^{-1}\b{W}$, where $\b{W}=\b{h}_{jw}\b{h}_{jw}^\dagger$ and, $\b{B}=\widetilde{\b{h}}\widetilde{\b{h}}^\dagger  +\sigma\|\b{c}\|^2\b{I}$ for, $\widetilde{\b{h}}=\b{H}_{jb}^\dagger\b{c}$.}
\end{theorem}
\begin{IEEEproof}
Assuming $\b{c}$ is fixed, the optimization problem in \eqref{equ-maximization problem of the jammer for general projection reduced to single direction} is reduced to,
\begin{equation*}
\begin{aligned}
&\max_{\b{v}} \frac{\b{h}_{jw}^\dagger \b{v} \b{v}^\dagger \b{h}_{jw}}{\b{c}^\dagger\left(\b{H}_{jb}  \b{v} \b{v}^\dagger\b{H}_{jb}^\dagger  +\sigma\b{I} \right)\b{c}}\\
&=\max_{\b{v}} \frac{\b{v}^\dagger \b{h}_{jw}\b{h}_{jw}^\dagger \b{v}}{\widetilde{\b{h}}^\dagger  \b{v} \b{v}^\dagger\widetilde{\b{h}} +\sigma\|\b{c}\|^2}\\
&=\max_{\b{v}} \frac{\b{v}^\dagger \b{h}_{jw}\b{h}_{jw}^\dagger \b{v}}{\b{v}^\dagger\left(\widetilde{\b{h}}\widetilde{\b{h}}^\dagger  +\sigma\|\b{c}\|^2\b{I}\right)\b{v}},
\end{aligned}
\end{equation*}
which is the Rayleigh quotient maximization problem \cite{mathai1992quadratic}. Note that we neglect the constants in \eqref{equ-maximization problem of the jammer for general projection reduced to single direction} since we are interested only in $\b{v}$. Thus, similarly to the proof of Theorem \ref{the-optimal solution for jammer covariance when helping Alice}, setting $\b{W}=\b{h}_{jw}\b{h}_{jw}^\dagger$ and, $\b{B}=\widetilde{\b{h}}\widetilde{\b{h}}^\dagger  +\sigma\|\b{c}\|^2\b{I}$, where $\widetilde{\b{h}}=\b{H}_{jb}^\dagger\b{c}$, the optimal $\b{v}$ is the eigenvector which corresponds to the highest eigenvalue
of the matrix $\b{B}^{-1}\b{W}$.
\end{IEEEproof}

\underline{\textbf{Fixed $\b{v}$:}} \newline
For any fixed $\b{v}$ we have the following theorem.
\begin{theorem}\label{the-optimal strategy for Bob for fixed v}
\emph{For a fixed $\b{v}$, the strategy for Bob, i.e., the solution to \eqref{equ-maximization problem of the jammer for general projection reduced to single direction} when the CSI of Willie is known, is the linear filter $\b{c}^*$ which is the eigenvector that corresponds to the highest eigenvalue of the matrix $\b{B}^{-1}\b{W}$, where $\b{W}=\b{h}_{ab}\b{h}_{ab}^\dagger$ and $\b{B}=\b{H}_{jb}  \b{v} \b{v}^\dagger\b{H}_{jb}^\dagger  +\sigma\b{I}$.}
\end{theorem}
\begin{IEEEproof}
Assuming $\b{v}$ is fixed the optimization problem in \eqref{equ-maximization problem of the jammer for general projection reduced to single direction} is reduced to,
\begin{equation*}
\begin{aligned}
&\max_{\b{c}} \frac{|\b{c}^\dagger\b{h}_{ab}|^2}{\b{c}^\dagger\left(\b{H}_{jb}  \b{v} \b{v}^\dagger\b{H}_{jb}^\dagger  +\sigma\b{I} \right)\b{c}}\\
&=\max_{\b{c}} \frac{\b{c}^\dagger\b{h}_{ab}\b{h}_{ab}^\dagger\b{c}}{\b{c}^\dagger\left(\b{H}_{jb}  \b{v} \b{v}^\dagger\b{H}_{jb}^\dagger  +\sigma\b{I} \right)\b{c}},
\end{aligned}
\end{equation*}
which is agin the Rayleigh quotient maximization problem \cite{mathai1992quadratic}. Thus, setting $\b{W}=\b{h}_{ab}\b{h}_{ab}^\dagger$ and $\b{B}=\b{H}_{jb}  \b{v} \b{v}^\dagger\b{H}_{jb}^\dagger  +\sigma\b{I}$, the optimal $\b{c}$ is the eigenvector which corresponds to the highest eigenvalue of the matrix $\b{B}^{-1}\b{W}$.
\end{IEEEproof}

As a consequence of Theorems \ref{the-optimal strategy for jammer for fixed c} and \ref{the-optimal strategy for Bob for fixed v} we now present the strategies in the following corollaries. The corollaries provide the specific linear filter $\b{c}$ and the direction vector $\b{v}$ of Bob and the jammer to be set in \eqref{equ-maximization problem of the jammer for general projection reduced to single direction}.
\begin{corollary}\label{cor-sub-optimal v for c=MRC}
\emph{When Bob employs a MRC for the channel between him and Alice while ignoring the AN, i.e., $\b{c}=\b{h}_{ab}$, the solution to \eqref{equ-maximization problem of the jammer for general projection reduced to single direction} is $\b{v}^*=\frac{\b{v}}{\|\b{v}\|}$, where $\b{v}$ is the eigenvector which corresponds to the highest eigenvalue of the matrix $\b{B}^{-1}\b{W}$, where $\b{W}=\b{h}_{jw}\b{h}_{jw}^\dagger$, $\b{B}=\widetilde{\b{h}}\widetilde{\b{h}}^\dagger  +\sigma\|\b{c}\|^2\b{I}$ and, $\widetilde{\b{h}}=\b{H}_{jb}^\dagger\b{c}$.}
\end{corollary}
\begin{corollary}\label{cor-sub-optimal c for v=to Willie}
\emph{When the jammer directs his AN towards Willie, i.e., $\b{v}=\frac{\b{h}_{jw}}{\|\b{h}_{jw}\|}$, the solution to \eqref{equ-maximization problem of the jammer for general projection reduced to single direction} is the linear filter $\b{c}^*$ which is the eigenvector that corresponds to the highest eigenvalue of the matrix $\b{B}^{-1}\b{W}$, where $\b{W}=\b{h}_{ab}\b{h}_{ab}^\dagger$ and $\b{B}=\b{H}_{jb}  \b{v} \b{v}^\dagger\b{H}_{jb}^\dagger  +\sigma\b{I}$.}
\end{corollary}

As mentioned above, the cancelation of the AN at Bob's receiver depends on the number of antennas that Bob and the jammer have. If Bob has more antennas than the jammer, i.e., $M>N$, then he can find a null-space for the columns of $\b{H}_{jb}$ such that the linear filter $\b{c}$ will satisfy $\b{c}^T \b{H}_{jb}=0$. 
In other words, Bob projects the received vector $\b{y}_b[i]$ at the $i$th channel use onto a subspace spanned by the null-space of $\b{H}_{jb}$. That is, we have

\begin{equation*}\label{equ-The received signals at Bob after null steering}
\begin{aligned}
\b{c}^T\b{y}_b[i]&=x[i]\b{c}^T\b{h}_{ab}+\b{c}^T\b{H}_{jb}\b{v}[i]+\b{c}^T\b{n}_b[i]\\
			&=x[i]\b{c}^T\b{h}_{ab}+\b{c}^T\b{n}_b[i].
\end{aligned}
\end{equation*}
The above implies that the vector $\widetilde{\b{h}}$ in Theorem \ref{the-optimal strategy for jammer for fixed c} is equal to zero. Note that Bob may choose $\b{c}$ to improve the SNR while still restricting $\b{c}$ to being in null-space of $\b{H}_{jb}$. 
\begin{figure*}[htbp]
   \centerline{
   \subfigure[$N=4,\ M=2$]{\includegraphics[width=0.33\textwidth]{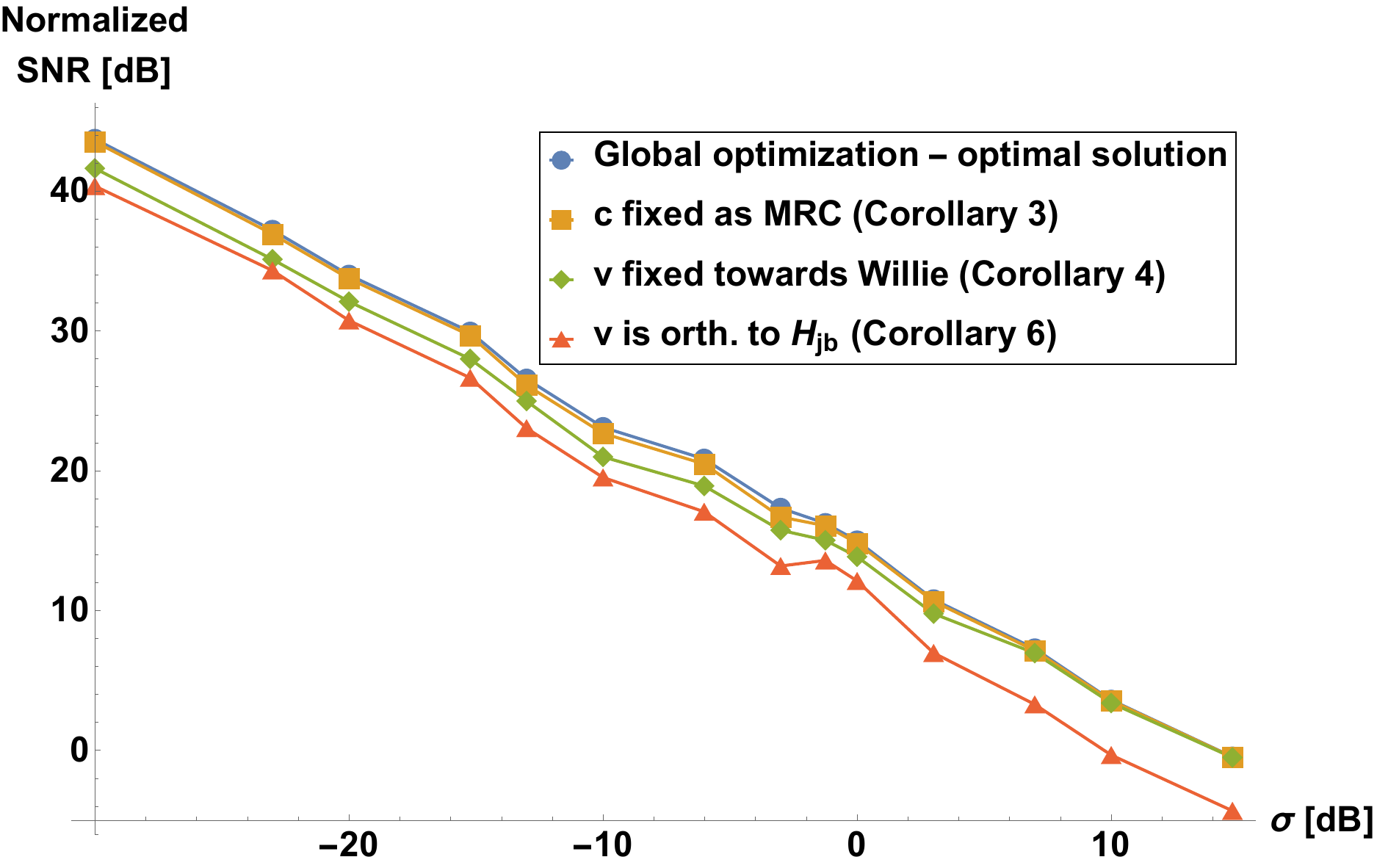}
       \label{fig-SNR_schemes_comparison_N_4_M_2}}
       \hfil
\subfigure[$N=M=3$]{\includegraphics[width=0.33\textwidth]{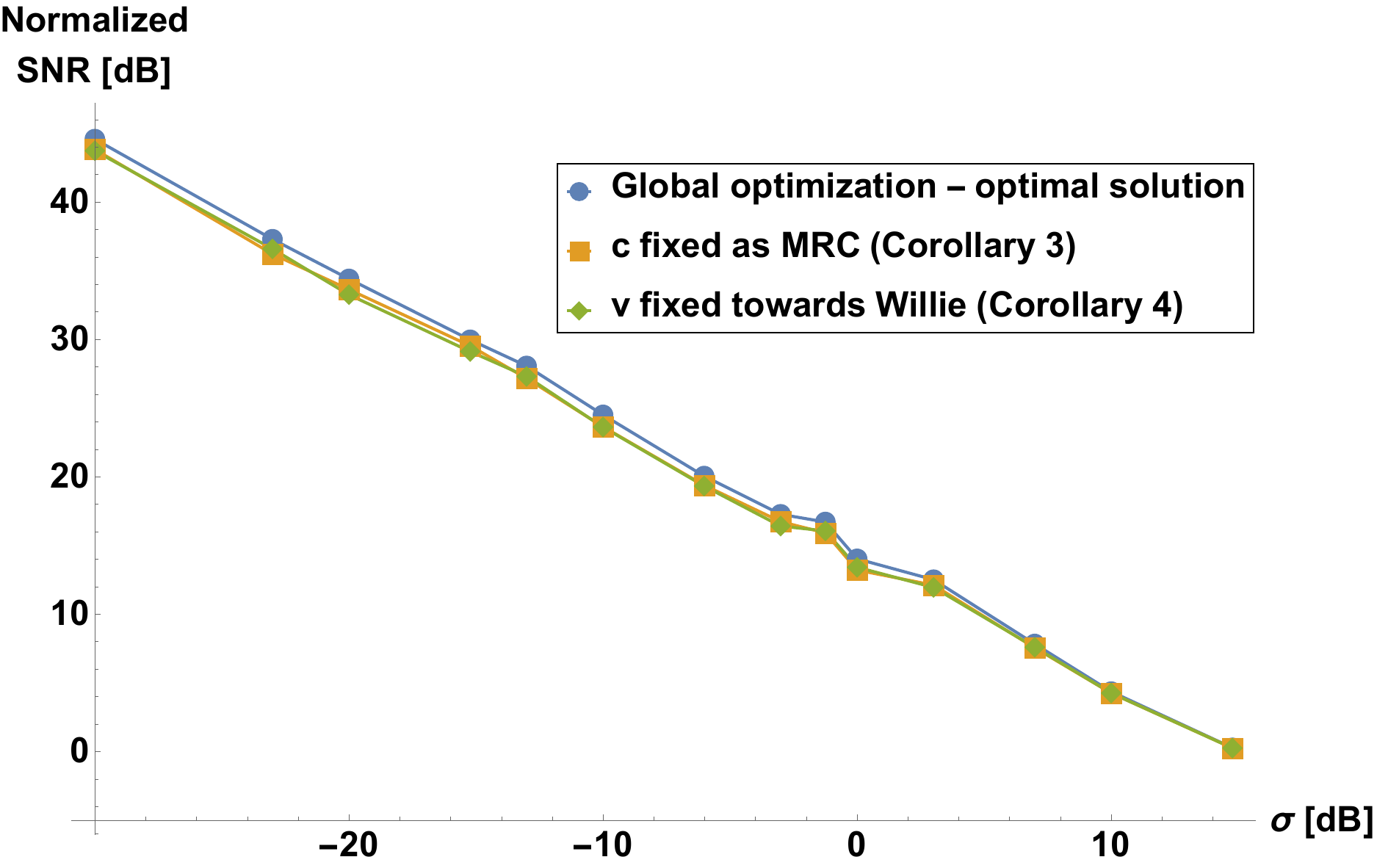}
       \label{fig-SNR_schemes_comparison_N_3_M_3}}
       \hfil
    \subfigure[$N=2, \ M=4$]{\includegraphics[width=0.33\textwidth]{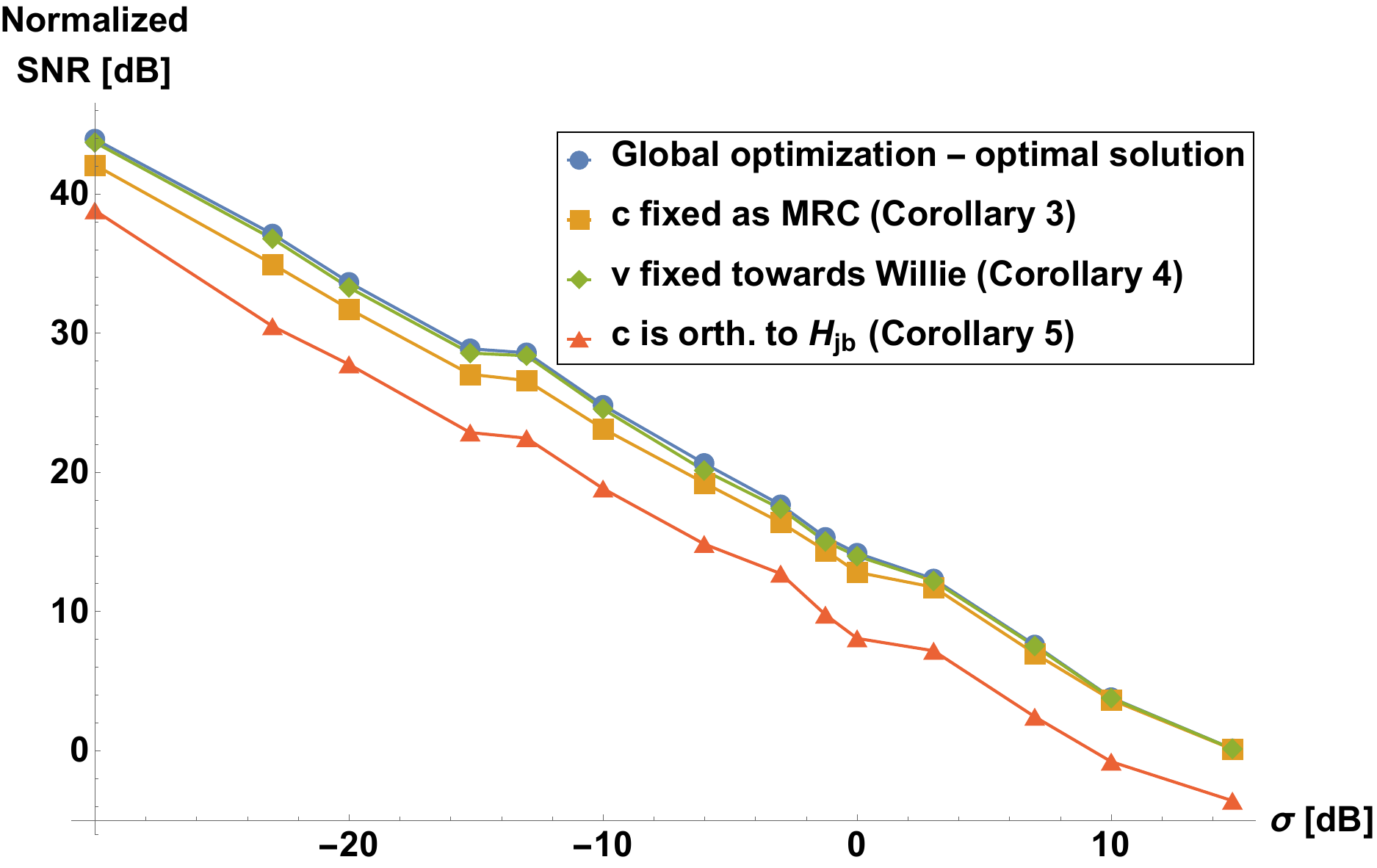}
       \label{fig-SNR_schemes_comparison_N_2_M_4}}}
   \caption{Simulation results which compare the optimal solution for the global optimization in Equation \eqref{equ-maximization problem of the jammer for general projection reduced to single direction} with the suggested sub-optimal schemes in Corollaries \ref{cor-sub-optimal v for c=MRC},\ref{cor-sub-optimal c for v=to Willie},\ref{cor-optimal solution for jammer covariance when M>1 and bob cancel AN} and \ref{cor-optimal solution for Bob linear filter M>1} as a function of $\sigma$. The three graphs are for a different constellations of the number of antennas. The SNR values, under $\epsilon$-covertness, are normalized so the leading constant coefficient is equal to 1.}
   \label{fig-SNR_schemes_comparison}
 \end{figure*}
Consequently, we have the following corollary.
\begin{corollary}\label{cor-optimal solution for jammer covariance when M>1 and bob cancel AN}
\emph{When Bob cancels the AN and chooses $\b{c}$ such that,
\begin{equation}\label{equ-MRC restricted to null-space Q}
 \b{c}=\argmax_{\b{c}\in nullspace(\b{H}_{jb})} \frac{|\b{c}^\dagger\b{h}_{ab}|^2}{\sigma_b^2\|\b{c}\|^2}.
\end{equation} 
The optimal strategy for the jammer is the following power allocation,
\begin{equation}\label{equ-optimal power allocation sigma}
\bs{\Sigma}=\b{v}^*{\b{v}^*}^\dagger,
\end{equation}
where, 
\begin{equation}\label{equ-optimal v direction}
\b{v}^*=\frac{\b{h}_{jw}}{\|\b{h}_{jw}\|}.
\end{equation}}
\end{corollary}
\begin{IEEEproof}
Since Bob cancels the received AN from the jammer, following the notations of Theorem \ref{the-optimal strategy for jammer for fixed c}, we have $\b{B}^{-1}=\b{B}=\sigma\|\b{c}\|^2\b{I}$ and $\b{W}=\b{h}_{jw}\b{h}_{jw}^\dagger$. Since the matrix $\b{BW}$ is of rank 1, it has only a single non-zero eigenvalue which $\b{h}_{jw}$ is its corresponding eigenvector.
\end{IEEEproof}

Corollary \ref{cor-optimal solution for jammer covariance when M>1 and bob cancel AN} essentially shows that when Bob can cancel the AN, the jammer can transmit his AN towards Willie without degrading the covert rate between Alice and Bob.

On the other hand, if the jammer has more antennas than Bob, i.e., $N>M$, then, the jammer may construct his covariance matrix such that the received AN power at Bob will fall first in the null-space of the columns of $\b{H}_{jb}^T$ and then as close as possible towards the direction of Willie. This strategy is different than the optimal one obtained in Theorem \ref{the-optimal solution for jammer covariance when helping Alice} since here the direction is always orthogonal to Bob. Thus, setting $\b{v}$ such that it satisfies $\b{H}_{jb}\b{v}=0$, the jammer is able to cancel the AN at Bob. Similarly, $\b{v}$ can be optimized further while restricted to be in the null-space.  Accordingly we have the following corollary.
\begin{corollary}\label{cor-optimal solution for Bob linear filter M>1}
\emph{When the jammer cancels the AN and chooses $\b{v}$ such that,
\begin{equation}\label{equ-To Willie restricted to null-space Q'}
\b{v}= \argmax_{\b{v}\in nullspace(\b{H}_{jb}^T)} \left(\b{h}_{jw}^\dagger \b{v} \b{v}^\dagger \b{h}_{jw}\right).
\end{equation}
The optimal strategy for Bob is to perform a MRC for the channel between Alice and himself. That is,
\begin{equation}\label{equ-optimal linear filter for Bob when jammer cancel AN}
\b{c}^*=\b{h}_{ab}.
\end{equation}}
\end{corollary}
\begin{IEEEproof}
Since the jammer cancels the AN in advance, following the notations of Theorem \ref{the-optimal strategy for Bob for fixed v}, we have $\b{B}^{-1}=\b{B}=\sigma\b{I}$ and $\b{W}=\b{h}_{ab}\b{h}_{ab}^\dagger$. Since the matrix $\b{BW}$ is of rank 1, it has only a single non-zero eigenvalue which $\b{h}_{ab}$ is its corresponding eigenvector (this solution is also known as the matched filter \cite{goldsmith2005wireless}).
\end{IEEEproof}
Corollary \ref{cor-optimal solution for Bob linear filter M>1} essentially states that once the jammer is able to cancel his AN in advance at Bob's receiver, Bob only performs maximal ratio combining to maximize his received SNR.  

Lastly, we have the following corollary which states that, as $\sigma$ grows (the low SNR regime), corollaries \ref{cor-sub-optimal v for c=MRC} and \ref{cor-sub-optimal c for v=to Willie} are asymptotically optimal. 
\begin{corollary}\label{cor-optimal solution for large sigma}
\emph{For $\sigma \rightarrow \infty$, the strategies suggested in corollaries \ref{cor-sub-optimal v for c=MRC} and \ref{cor-sub-optimal c for v=to Willie} are asymptotically optimal in the sense of approching the global optimum of \eqref{equ-maximization problem of the jammer for general projection reduced to single direction}.}
\end{corollary}
\begin{IEEEproof}
At the limit of $\sigma \rightarrow \infty$, the effect of the AN noise term in the denominator of the SNR expression in \eqref{equ-maximization problem of the jammer for general projection reduced to single direction} becomes negligible thus \eqref{equ-maximization problem of the jammer for general projection reduced to single direction} is reduced to
\begin{equation}
\max_{\b{v,c}} \frac{\left(\b{h}_{jw}^\dagger \b{v} \b{v}^\dagger \b{h}_{jw}\right)|\b{c}^\dagger\b{h}_{ab}|^2}{\sigma\|\b{c}\|^2},
\end{equation}
which is maximized when Bob employ an MRC and the AN noise transmission is in the direction of Willie.
\end{IEEEproof}

\begin{remark}[\emph{The case $M=N$}]
In case the jammer and Bob have the same number of antennas, the sub-optimal schemes suggested herein are not applicable due to the fact that $\b{H}_{jb}$ and $\b{H}_{jb}^\dagger$ are square matrices and thus a null-space does not exist for them. Therefore, other types of schemes must be performed. We note that in any case, one can use either Corollary \ref{cor-sub-optimal v for c=MRC} or Corollary \ref{cor-sub-optimal c for v=to Willie} as a possible scheme.
\end{remark}

Simulation results which compare the normalized average SNR value, as a function of $\sigma$, are depicted in Figure \ref{fig-SNR_schemes_comparison} for the cases of $M<N$, $M=N$ and $M>N$. The figures compare \emph{the optimal solution for the global optimization problem \eqref{equ-maximization problem of the jammer for general projection reduced to single direction}} (solved using Mathematica 12) with the suggested sub-optimal schemes given in Corollaries \ref{cor-sub-optimal v for c=MRC},\ref{cor-sub-optimal c for v=to Willie},\ref{cor-optimal solution for jammer covariance when M>1 and bob cancel AN} and \ref{cor-optimal solution for Bob linear filter M>1}. 

We note that the strategy of canceling the AN given in Corollaries \ref{cor-optimal solution for jammer covariance when M>1 and bob cancel AN} and \ref{cor-optimal solution for Bob linear filter M>1} are depicted in Figures \ref{fig-SNR_schemes_comparison_N_4_M_2} and \ref{fig-SNR_schemes_comparison_N_2_M_4} for the cases of $M<N$ and $M>N$, respectively. When $M=N$ neither Bob nor the jammer can cancel the AN as the matrices $\b{H}_{jb}$ and $\b{H}_{jb}^T$ are square. Thus, only corollaries \ref{cor-sub-optimal v for c=MRC} and \ref{cor-sub-optimal c for v=to Willie} are depicted in Figure \ref{fig-SNR_schemes_comparison_N_3_M_3}.


One can observe several interesting insights from the simulation results. The first is that when $M<N$, choosing the strategy that fixes $\b{c}$ to be the MRC and optimizing only on $\b{v}$ (Corollary \ref{cor-sub-optimal v for c=MRC}) performs very well compared to the optimal solution of \ref{equ-maximization problem of the jammer for general projection reduced to single direction}. On the other hand, when $M>N$, it is better to choose the strategy which fixes $\b{v}$ to be in the direction of Willie and optimize only on $\b{c}$ (Corollary \ref{cor-sub-optimal c for v=to Willie}). This can be explained by the number of DoF which are left for the optimization. That is, for example, when $M<N$ and we fix $\b{c}$ we are left with $N$ variables in the optimization problem which can provide a more accurate search for the optimum. A piece of additional evidence for that is the fact that the two curves unite when $M=N$. 

The second observation is that as $\sigma$ grows (low SNR scenario), consistent to Corollary \ref{cor-optimal solution for large sigma}, the sub-optimal schemes, with the exception of the AN cancelation scheme, coincide with the optimal SNR making them asymptotically optimal schemes. 

Finally, one can observe that the strategy of canceling the AN is always worse than MRC or directing the noise towards Willie. The reason lies in the fact that once the AN is canceled, the inner products in the numerator are restricted to be only in the specific null-space of Bob (as can be seen in equations \eqref{equ-MRC restricted to null-space Q} and \eqref{equ-To Willie restricted to null-space Q'}). This evidently, restricts much the SNR value as the figures depict. Moreover, even when $\sigma\rightarrow\infty$ this restriction remains and thus there is no convergence to the optimal SNR. 

\subsection{Unknown CSI of Willie}

When the CSI of Willie is unknown, similar to the $M=1$ case, Alice transmits with the power given in Lemma \ref{lem-Alice power for covert system with no CSI} in order to guarantee a covert transmission and the jammer transmits with all of his available power equally divided to $d$ directions. Accordingly, we have the following global optimization problem for Bob and the jammer,

\begin{equation}\label{equ-maximization problem of the jammer for general projection No CSI}
\begin{aligned}
&\max_{\b{V},d,\b{c}} \frac{\frac{\epsilon P_{max}}{12 e \ln{\frac{6}{\epsilon}}}\left(\frac{\epsilon}{6}\right)^{\frac{1}{d}}|\b{c}^\dagger\b{h}_{ab}|^2}{\b{c}^\dagger\left(p_j\b{H}_{jb} \b{V} \b{X} \b{V}^\dagger\b{H}_{jb}^\dagger  +\sigma_b^2\b{I} \right)\b{c}}\\
& s.t. \quad \xi_l = \frac{1}{d},\ \ l=1,...,d.
\end{aligned}
\end{equation} 

Again, to avoid the complexity in \eqref{equ-maximization problem of the jammer for general projection No CSI}, we restrict the attention to several sub-optimal solutions. Specifically, in case Bob has $M>N$ antennas, he can cancel the AN by projecting it onto the null-space of $\b{H}_{jb}$ and choosing $\b{c}$ which is closest to $\b{h}_{ab}$. The jammer's strategy for this case is given in the following theorem.

\begin{theorem}\label{the-optimal solution for jammer covariance when M>1 and bob cancel AN No CSI Willie}
\emph{Assume $M>N$ and set:
\begin{equation}
 \b{c}^*=\argmax_{\b{c}\in nullspace(\b{H}_{jb})} \frac{|\b{c}^\dagger\b{h}_{ab}|^2}{\sigma_b^2\|\b{c}\|^2}.
\end{equation}
Then the optimal strategy for the jammer when the jammer has no CSI of Willie and he transmits with all of his available power, $p_j$, equally divided to all directions, is the following power allocation,
\begin{equation}\label{equ-optimal power allocation sigma when M>1 and bob cancel AN No CSI Willie}
\bs{\Sigma}= \frac{1}{N}\b{I}_{N\times N}.
\end{equation}}
\end{theorem}
\begin{IEEEproof}
Once the interference term in the denominator of \eqref{equ-maximization problem of the jammer for general projection No CSI} is gone, the jammer should maximize the number of direction $d$ which is $N$ regardless of the specific directions.
\end{IEEEproof}
The above theorem implies that once Bob can cancel the noise, the jammer should try to impair Willie's detection as much as possible. This is different from the result in Theorem \ref{the-optimal solution for jammer covariance when helping Alice no CSI} since now Bob takes an active part in the reception which enables the jammer to concentrate on Willie and to transmit in every direction (as Willie's channel is unknown).
 
In case the jammer has more antennas than Bob ($M<N$), a naive solution would be to cancel the AN by setting the matrix $\b{V}$ to span the null-space of $\b{H}_{jb}^T$. In this case the rank of $\b{V}$ is $N-M$, i.e., $d=N-M$, which affects the numerator in \eqref{equ-maximization problem of the jammer for general projection No CSI} directly and especially when $N$ is on the same order as $M$. However, according to Corollary \ref{cor-optimal solution for Bob linear filter M>1}, we know that in case the jammer can completely remove the AN interference, Bob's optimal linear filter is the MRC $\b{c}^*$ as given in the corollary. In fact, since Bob uses a linear filter cancelling his AN should require only one DoF. We thus suggest a scheme for which the jammer relies upon this choice of Bob for the linear filter $\b{c}^*$ and can null the equivalent one dimensional channel and benefiting from the remaining DoF he has. This strategy is given in the following theorem. 
\begin{theorem}\label{the-optimal solution for jammer covariance when M>1 and bob uses MRC No CSI Willie}
\textit{For the fixed linear filter $\b{c}^*=\b{h}_{ab}$, the optimal strategy for the jammer when the CSI of Willie is unknown and the jammer transmits AN with covariance matrix $\bs{\Sigma}$ such that his total transmission power, $p_j$ , is equally
divided into $d$ arbitrary directions, is the following power allocation,
\begin{equation}
\bs{\Sigma}= \b{V}^*\b{X}^*{\b{V}^*}^\dagger,
\end{equation}
where in case $\sigma=\frac{\sigma_b^2}{p_j}<\frac{\|\widetilde{\b{h}}\|^2}{N\left( {\frac{\epsilon}{6}}^{-\frac{1}{N(N-1)}}-1 \right)\| \b{c}^* \|^2}$ ,
\begin{equation}
\begin{aligned}
&\b{V}^*=\widetilde{\b{Q}} \ \text{ and, } 
&\b{X}^*=\frac{1}{N-1}I_{(N-1)\times(N-1)}.
\end{aligned}
\end{equation}
Otherwise,
\begin{equation}
\begin{aligned}
&\b{V}^*=\left(\widetilde{\b{Q}} \ , \ \frac{\widetilde{\b{h}}}{\|\widetilde{\b{h}}\|} \right)  \text{  and, } 
&\b{X}^*=\frac{1}{N}I_{N\times N},
\end{aligned}
\end{equation}
where $\widetilde{\b{h}}=\b{H}_{jb}^\dagger \b{c}^*$ and $\widetilde{\b{Q}}$ is a matrix whose columns span the null-space of $\widetilde{\b{h}}$.}
\end{theorem}

\begin{IEEEproof}
Considering that $\b{c}^*=\b{h}_{ab}$ is fixed and using notations from the proof of Theorem \ref{the-optimal solution for jammer covariance when helping Alice no CSI}, the maximization problem \eqref{equ-maximization problem of the jammer for general projection No CSI} can be reduced to,
\begin{equation*}
\begin{aligned}
&\max_{\b{V},d} \frac{\left(\epsilon'\right)^{\frac{1}{d}}}{{\b{c}^*}^\dagger\b{H}_{jb} \b{V} \b{X} \b{V}^\dagger\b{H}_{jb}^\dagger \b{c}^*  +\sigma \| \b{c}^* \|^2}\\
&= \max_{\b{V},d} \frac{\left(\epsilon'\right)^{\frac{1}{d}}}{\widetilde{\b{h}}^\dagger \b{V} \b{X} \b{V}^\dagger \widetilde{\b{h}}  +\sigma \| \b{c}^* \|^2},\\
\end{aligned}
\end{equation*}
where $\widetilde{\b{h}}=\b{H}_{jb}^\dagger \b{c}^*$. The above can be simplified further to
\begin{equation*}
 \max_{d,\b{b}} \frac{ \left(\epsilon'\right)^{\frac{1}{d}} }{\sum_{l=1}^d \frac{1}{d} b_l^2+\sigma \| \b{c}^* \|^2}
\end{equation*}
where $\b{b}=\b{V}^\dagger \widetilde{\b{h}}$. We result with a similar to the expression in the proof of Theorem \ref{the-optimal solution for jammer covariance when helping Alice no CSI} and thus the rest follows similarly.
\end{IEEEproof}

\section{Conclusion}
In this work, we considered the problem of covert communication in the presence of a jammer who is equipped with multiple antennas. The jammer assists Alice and Bob by transmitting AN that creates uncertainty at Willie's detector and enables Alice and Bob to communicate with a positive rate. The transmission strategies which affect this rate are examined and analyzed in the form of optimization problems for the cases of full CSI and partial CSI at the jammer. The use of multiple antennas provides a multiplicative gain to the covertness achieving transmission power with respect to the case of a single antenna jammer. Specifically, under certain assumptions, this gain is an increasing function of the number of antennas at the jammer. 

When Bob is equipped with a single antenna the optimal strategy for a jammer with full CSI is beamforming the AN with all available power to a specific direction, which is close to the direction of Willie while being orthogonal to Bob as much as possible. On the other hand, in the case of partial CSI, when the jammer does not know the channel toward Willie, the optimal AN transmission strategy is to transmit isotropically to all directions or to the null-space of Bob, where the choice depends on certain channel conditions. 
Finally, when Bob is equipped with multiple antennas and is assumed to employ a linear receiver, several sub-optimal schemes are suggested along with numerical results. These results show that the strategy of letting the jammer transmit his AN towards Willie while letting Bob employ MRC is an asymptotically optimal solution for the global optimization problem which includes both Willie's and Bob's strategies. 

\section*{Acknowledgment}
The authors would like to thank Alejandro Cohen for discussions that helped prompt this work.

\appendices

\bibliographystyle{IEEEtran}
\bibliography{Bibliography_cc}

\end{document}